\documentclass[10pt]{article}
\usepackage{amsmath, amssymb,amsfonts,bbm,verbatim,multirow}
\usepackage{epic,eepic,psfrag,epsfig}
\usepackage[sf,bf,SF,footnotesize]{subfigure}
\usepackage{graphicx}
\usepackage{amsthm}
\usepackage{amscd}
\usepackage{epsfig}
\usepackage{fullpage}
\usepackage{natbib}%{natbibspacing}
\usepackage{setspace}
\usepackage{enumerate}
\usepackage{array}
\usepackage{pifont}
\usepackage[small]{caption}
\newcommand{\cmark}{\ding{51}}
\newcommand{\xmark}{\ding{55}}
\newcommand{\specialcell}[2][c]{\begin{tabular}[#1]{@{}c@{}}#2\end{tabular}}
\def\mathbi#1{\textbf{\em #1}}
\def\argmax{\mathop{\mathrm{arg\,max}}}
\newtheorem{thm}{Theorem}[section]

\newtheorem{lem}[thm]{Lemma}

\newtheorem{cor}[thm]{Corollary}
\numberwithin{equation}{section}
\newenvironment{prooftitle}[1]{{\noindent \textsc{Proof #1}}\\}

\begin{document}

\title{Semiparametric Time Series Models with Log-concave Innovations: \\Maximum Likelihood Estimation and its Consistency}
\author{Yining Chen \\
  Statistical Laboratory\\
  University of Cambridge\\
  Y.Chen@statslab.cam.ac.uk
}

\maketitle

\onehalfspacing

\begin{abstract}
  \noindent
We study semiparametric time series models with innovations following a log-concave distribution. We propose a general maximum likelihood framework which allows us to estimate simultaneously the parameters of the model and the density of the innovations. This framework can be easily adapted to many well-known models, including ARMA, GARCH and ARMA-GARCH. Furthermore, we show that the estimator under our new framework is consistent in both ARMA and ARMA-GARCH settings. We demonstrate its finite sample performance via a thorough simulation study and apply it to model the daily log-return of FTSE 100 index and the rabbit population.
  \bigskip


\end{abstract}

\section{Introduction}
\label{Sec:intro}
Statistical analysis of time series is an important issue in many areas of science. Many existing time series models postulate Gaussian innovations. Statistical inference is then typically based on the idea of maximum likelihood estimation. Some well-known examples include the autoregressive moving average (ARMA) models \citep{BrockwellDavis1991} and the generalized autoregressive conditionally heteroscedastic (GARCH) models \citep{Bollerslev1986}. However, it is known that time series with non-Gaussian innovations frequently occur in health, social and environmental sciences \citep{DLZ2002}. Often, the Gaussian quasi-maximum likelihood estimator (GQMLE) is used to alleviate this issue, and in most circumstances, the resulting estimates are still consistent (cf. \citet{FrancqZakoian2004}). Nevertheless, we argue that there are circumstances where semiparametric models are preferable, because estimating the distribution function of the innovations enhances our understanding of the data. For example, utilizing its quantiles can lead to more informative predictions \citep{KoenkerHallock2001}.

As an early attempt to model the innovation density nonparametrically, \citet{EGR1991} proposed a semiparametric autoregressive conditionally heteroscedastic (ARCH) model based on a nonparametric density estimation technique called discrete maximum penalized likelihood estimation. \citet{DKW1997} suggested an adaptive estimator (AE) for ARMA based on the kernel density estimator. See \citet{Kreiss1987}, \citet{DrostKlaassen1997}, \citet{SunStengos2006} and \citet{LingMcAleer2003} for related work on other time series models.  However, we argue that the above-mentioned estimators may potentially suffer from the following drawbacks:
\begin{enumerate}[(a)]
\setlength{\itemsep}{0pt}
\setlength{\parskip}{0pt}
\setlength{\parsep}{0pt}
\item they mainly focus on estimating the parametric part of the models;
\item their finite-sample performances depend heavily on the choice of tuning parameters, especially when the sample size is not too large. However, none of the above-cited work gives practical guidelines on how to set tuning parameters;
\item often some restrictive conditions are imposed, for instance, it is generally assumed that the innovation distribution has a continuous density function. Furthermore, both \citet{Kreiss1987} and \citet{LingMcAleer2003} require the density function of the innovations to be symmetric.
\end{enumerate}

Motivated by recent developments in shape-constrained density estimation, in this paper we take a different approach by assuming that the innovations have a \emph{log-concave} density (i.e. the logarithm of the density function is concave). The class of log-concave densities contains many commonly encountered parametric families of univariate distributions, including normal, gamma with shape parameter at least 1, Weibull distributions with shape parameter at least 1, beta($\alpha,\beta$) with $\alpha,\beta \ge 1$, logistic, Laplace (double exponential) and Gumbel; see \citet{BagnoliBergstrom2005} for more examples. Throughout this paper, we denote the class of log-concave densities by $\mathcal{F}$.

Our new modeling framework is as follows. Denote a class of separated semiparametric time series models by $(f, \boldsymbol{\theta})$, where $f$ is the density function of the independent and identically distributed (\emph{i.i.d.}) innovations, and $\boldsymbol{\theta}$ is the parameter vector taking values in a parameter space $\Theta$. Let $l(f, \boldsymbol{\theta})$ be its log-likelihood function. Denote the true density of the innovations and the true value of parameter vector by $f_0$ and $\boldsymbol{\theta}_0$ respectively. We propose to estimate $f_0$ and $\boldsymbol{\theta}_0$ by
\[
	(\hat{f},\hat{\boldsymbol{\theta}}) \in  \argmax_{f \in \mathcal{F},\, \boldsymbol{\theta} \in \Theta} \, l(f, \boldsymbol{\theta}).
\]
We call $(\hat{f}, \hat{\boldsymbol{\theta}})$ the \emph{log-concave maximum likelihood estimator} (LCMLE). 

Our method can be viewed as a generalization of \citet{DSS2011}, where this type of estimators was first proposed and studied for the linear regression models. It is also related to sieved estimators such as those in \citet{CLS2012}. The main advantages of our method include the following:
\begin{enumerate}[(a)]
\setlength{\itemsep}{0pt}
\setlength{\parskip}{0pt}
\setlength{\parsep}{0pt}
\item it is \emph{free} of tuning parameters;
\item it simultaneously estimates the density function of the innovations and the parametric part of the model;
\item it is straightforward to implement;
\item it is easy to adapt to a wide class of time series models with only minor modifications;
\item for many classes of models, if $f_0$ is log-concave, then both $\hat{f}$ and $\hat{\boldsymbol{\theta}}$ are consistent;
\item even if $f_0$ is not log-concave, under weak assumptions (mainly the finite first moment of $f_0$), $\hat{\boldsymbol{\theta}}$ can still be a consistent estimator of $\boldsymbol{\theta}_0$;
\item it offers huge potential improvement over both the GQMLE and the AE in terms of finite sample performance.
\end{enumerate}

Here we list some applicable areas for our procedure. We argue that our approach gives an alternative to many of the statistical models listed below.
\begin{enumerate}[(a)]
\setlength{\itemsep}{0pt}
\setlength{\parskip}{0pt}
\setlength{\parsep}{0pt}
\item \underline{Streamflow and other hydrological data}: Investigations \citep{TYK1976} show that the independent residuals of autoregressive daily flow models have distributions whose tails are not heavier than exponential. \citet{DamslethElShaarawi1989} studied the ARMA models with Laplace innovations and used it to model the sulphate concentration in lakes in Ontario, Canada. 
\item \underline{Animal populations}: \citet{LiMcLeod1988} studied the ARMA models with skewed innovations, and fitted an autoregressive model with gamma innovations to the Canadian lynx data. See Section~\ref{Sec:rabbit} for an empirical example.
\item \underline{Financial data}: The GARCH model with Laplace innovations was shown to be superior to that with Gaussian innovations by \citet{GrangerDing1995} for the S\&P 500 index. In addition, \citet{HMP2006} reported that the GARCH model with innovations being the convolution of Laplace and Gaussian (which is log-concave) offers a plausible description of the daily stock return series in Germany. Recently, \citet{TZA2010} studied the ARMA-GARCH models with asymmetric Laplace innovations and applied them to model real estate returns. See also Section~\ref{Sec:FTSE} for a real data example.
\end{enumerate}

The nonparametric log-concave maximum likelihood density estimator was studied in the \emph{i.i.d.} setting by \citet{Walther2002}, \citet{PWM2007}, \citet{DumbgenRufibach2009}, \citet{BRW2009}, \citet{CSS2010}, \citet{CuleSamworth2010}, \citet{SHD2011} and \citet{DHR2011}. These references contain characterizations of the estimator, asymptotics and algorithms for its computation. Regarding its applications, see \citet{DSS2011}, \citet{Rufibach2012} and \citet{SamworthYuan2012}, where it has been applied to the isotonic / linear regression, the receiver operating characteristic (ROC) curve estimation and independent component analysis. Yet, to the best of our knowledge, none of the existing work concerns dependent data structures such as the stochastic processes studied in this paper. In fact, this paper gives very positive answers to the questions raised recently by \citet{XiaTong2010} and \citet{Yao2010}. For other popular shape constraints, one may refer to \citet{GJW2001}, \citet{SereginWellner2010} and \citet{KoenkerMizera2010}. 

The rest of the paper is organized as follows. In Section~\ref{Sec:linear}, we apply our method to the class of ARMA models. We display in detail how the LCMLE is constructed in Section~\ref{Sec:linearconstruct}. Theoretical results regarding its existence and consistency are given in Section~\ref{Sec:lineartheory}. A variant of the LCMLE is suggested in Section~\ref{Sec:linearsmooth}, which offers further potential improvement in small sample sizes and provides a nice link to the smoothed log-concave maximum likelihood estimator studied by \citet{DumbgenRufibach2009} and \citet{ChenSamworth2013}.  

Section~\ref{Sec:nonlinear} adapts the framework to a particular nonlinear setting, where ARMA-GARCH models are considered. The challenge of constructing the LCMLE is taken up in Section \ref{Sec:nonlinearconstruct}, while results concerning its existence and consistency are described in Section \ref{Sec:nonlineartheory}. It is worth noting that in Sections~\ref{Sec:lineartheory} and \ref{Sec:nonlineartheory}, our theory is developed under both correct and incorrect model specification of the innovation distribution. 

Section~\ref{Sec:computing} is devoted to the computation of the LCMLE. Simulation studies follow in Section~\ref{Sec:simgq} and \ref{Sec:simae}, confirming the significantly improved finite sample performance over the GQMLE and the AE in the setting of non-Gaussian innovations. Moreover, we demonstrate that even in the case where the innovations are Gaussian, the performance of our LCMLE remains comparable to that of its competitors. These simulation results show great promise of the LCMLE, even though its asymptotic distributional theory remains to be investigated further. %Finally, Section~\ref{Sec:FTSE} gives an application of our methodology to the daily log-return of FTSE 100 index. 

Finally, Section~\ref{Sec:realdata} gives applications of our methodology to model the daily log-return of FTSE 100 index and the Yorkshire rabbit (\emph{Oryctolagus cuniculus}) population. We defer all proofs to the appendix.    

%Many proofs and technical arguments are deferred to the appendix. A longer and more detailed version of this paper is \citet{Chen2013}. It contains all the proofs and additional examples.

\section{ARMA models}
\label{Sec:linear}
In this section, we consider the ARMA($p,q$) process with observations $\{X_t\}$. The model is defined as 
\[
	X_t = \sum_{i=1}^p a_i X_{t-i} + \sum_{i=1}^q b_i \epsilon_{t-i} + \epsilon_t, 
\]
where $\{\epsilon_t\}$ are \emph{i.i.d.} random variables, and where $a_1,\ldots,a_p,b_1,\ldots,b_q$ are real coefficients. 

Arguably, ARMA models are the most popular linear models used by time series practitioners. See \citet{BrockwellDavis1991} for a thorough survey of the background. Our goal in this section is to estimate the parameters $a_1, \ldots, a_p,b_1,\ldots,b_q$ and the distribution of $\{\epsilon_t\}$ simultaneously.

\subsection{The log-concave maximum likelihood estimator}
\label{Sec:linearconstruct}
Assume that the observations $X_1,\ldots,X_n$ are from an ARMA($p,q$) process, where the orders $p$ and $q$ are known. The vector of the parameters
\[
	\boldsymbol{\theta} = (\mathbi{a}^T,\mathbi{b}^T)^T = (a_1, \ldots, a_p,b_1,\ldots,b_q)^T
\]
belongs to a parameter space $\Theta \subseteq \mathbb{R}^{p+q}$.

Let $\boldsymbol{\theta}_0 = (\mathbi{a}_0^T,\mathbi{b}_0^T)^T = (a_{01}, \ldots, a_{0p},b_{01},\ldots,b_{0q})^T$ and $Q_0$ denote respectively the true value of the parameter vector and the true distribution of the innovations. 

Let $\Phi$ be the family of concave functions $\phi: \mathbb{R} \rightarrow [-\infty,\infty)$ which are upper semicontinuous and coercive in the sense that $\phi(x) \rightarrow -\infty$ as $| x | \rightarrow \infty$. Furthermore, denote the set of concave log-densities by
\[
	 \Phi_0 = \left\{\phi \in \Phi : \int e^{\phi(x)} dx = 1 \right\}.
\]

The following conditions are imposed to construct the LCMLE:
\begin{enumerate}
\setlength{\itemsep}{0pt}
\setlength{\parskip}{0pt}
\setlength{\parsep}{0pt}
\item[\textbf{(A.1)}] $Q_0$ is a distribution with density function $f_0$ and has finite expectation; 
\item[\textbf{(A.2)}] $\boldsymbol{\theta}_0 \in \Theta$, where $\Theta$ is closed;
\item[\textbf{(A.3)}] $\Theta$ is a bounded subset of $\mathbb{R}^{p+q}$.
\end{enumerate}
 
The log-concave log-likelihood can be expressed as 
\[
	l_n(\phi, \boldsymbol{\theta}) = l_n(\phi, \boldsymbol{\theta}; X_1, \ldots, X_n) = \frac{1}{n}\sum_{t=1}^n \phi (\tilde{\epsilon}_t(\boldsymbol{\theta})),
\]
where $\phi \in \Phi_0$, $\boldsymbol{\theta} \in \Theta$ and $\{\tilde{\epsilon}_t(\boldsymbol{\theta})\}$ are the  estimated innovations computed recursively by
\[
	\tilde{\epsilon}_t(\boldsymbol{\theta}) = X_t - \sum_{i=1}^p a_i X_{t-i} - \sum_{i=1}^q b_i \tilde{\epsilon}_{t-i}(\boldsymbol{\theta}), \mbox{ for } t = 1,\ldots,n.
\]
The choice of the unknown initial values $X_0, \ldots,X_{1-p},\tilde{\epsilon}_0(\boldsymbol{\theta}), \ldots, \tilde{\epsilon}_{1-q}(\boldsymbol{\theta})$ can be shown to be unimportant asymptotically (see appendix for details). For simplicity, these initial values are taken to be fixed (i.e. neither random nor functions of the parameters).

Intuitively, one would seek to maximize $l_n(\phi, \boldsymbol{\theta})$ over $\Phi_0 \times \Theta$. However, it turns out that this naive optimization approach is very computationally intensive. We therefore employ the standard trick of \citet{Silverman1982} and propose the following procedure:

\begin{enumerate}
\setlength{\itemsep}{0pt}
\setlength{\parskip}{0pt}
\setlength{\parsep}{0pt}
\item[(i)] Let $(\hat{\phi}_n,\hat{\boldsymbol{\theta}}_n)$ be a maximizer of 
\begin{align}
\label{Eq:armamle1}
	\Lambda_n(\phi, \boldsymbol{\theta}) = \Lambda_n(\phi, \boldsymbol{\theta}; X_1, \ldots, X_n) = \frac{1}{n} \sum_{t=1}^n \phi (\tilde{\epsilon}_t(\boldsymbol{\theta})) - \int e^{\phi(x)}dx + 1
\end{align}
over all $(\phi, \boldsymbol{\theta}) \in \Phi \times \Theta$.
\item[(ii)] Return
\begin{align}
\label{Eq:armamle2}
	\hat{f}_n(x) = e^{\hat{\phi}_n(x)} \quad  \mbox{ and } \quad  \hat{\boldsymbol{\theta}}_n, 
\end{align}
where we call $\hat{f}_n$ and $\hat{\boldsymbol{\theta}}_n$ respectively the LCMLE of $f_0$ and $\boldsymbol{\theta}_0$ in ARMA. 
\end{enumerate}

\noindent\textbf{Remark}: For any fixed $\boldsymbol{\theta}$, the maximizer $\phi_{\boldsymbol{\theta}} = \argmax_{\phi \in \Phi}\Lambda_n(\phi,\boldsymbol{\theta})$ automatically satisfies $\int e^{\phi_{\boldsymbol{\theta}}(x)}dx=1$. Therefore, $e^{\hat{\phi}_n(x)}$ always defines a density.

\subsection{Theoretical properties}
\label{Sec:lineartheory}
\begin{thm}[Existence in ARMA]
\label{Thm:armaexist}
For every $n > p+q+1$, under assumptions \textbf{(A.1)} -- \textbf{(A.3)}, the LCMLE $(\hat{f}_n, \hat{\boldsymbol{\theta}}_n)$ defined in (\ref{Eq:armamle2}) exists with probability one.
\end{thm}

In the case $q=0$ (autoregressive models), assumption \textbf{(A.3)} is not needed to guarantee the existence of the LCMLE. In particular, as is justified by the following corollary, one can just take $\Theta = \mathbb{R}^p$.
\begin{cor}
\label{Cor:arexist}
If $q = 0$, then for every $n > p+1$, under assumptions \textbf{(A.1)} -- \textbf{(A.2)}, the LCMLE $(\hat{f}_n, \hat{\boldsymbol{\theta}}_n)$ defined in (\ref{Eq:armamle2}) exists with probability one.
\end{cor}

Define the ARMA polynomials as follows:
\begin{align}
	\mathbi{A}_{\boldsymbol{\theta}}(z) = 1 - \sum_{i=1}^p a_i z^i \quad \mbox{ and } \quad \mathbi{B}_{\boldsymbol{\theta}}(z) = 1 + \sum_{i=1}^q b_i z^i.
\end{align}
To establish the consistency of the LCMLE, we impose two more assumptions:
\begin{enumerate}
\setlength{\itemsep}{0pt}
\setlength{\parskip}{0pt}
\setlength{\parsep}{0pt}
\item[\textbf{(A.4)}] For all $\boldsymbol{\theta} \in \Theta$, $\mathbi{A}_{\boldsymbol{\theta}}(z)\mathbi{B}_{\boldsymbol{\theta}}(z) \ne 0$ for all $z \in \mathbb{C}$ such that $|z| \le 1$;
\item[\textbf{(A.5)}] If $p>0$ and $q>0$, $\mathbi{A}_{\boldsymbol{\theta}_0}(z)$ and $\mathbi{B}_{\boldsymbol{\theta}_0}(z)$ have no common roots and $|a_{0p}| + |b_{0q}| \ne 0$.
\end{enumerate}

\noindent\textbf{Remarks}: 
\begin{enumerate}[1.]
\setlength{\itemsep}{0pt}
\setlength{\parskip}{0pt}
\setlength{\parsep}{0pt}
\item Under assumption \textbf{(A.4)}, it can be shown in the spirit of Proposition~13.3.2 of \citet{BrockwellDavis1991} that observations $\{X_t\}$ are drawn from a strictly stationary and ergodic process. It also restricts our attention to causal and invertible ARMA processes.
\item The ARMA models without assumption \textbf{(A.5)} are not identifiable. Assumption \textbf{(A.5)} also allows for an overidentification of either $p$ or $q$, but not both.
\end{enumerate}

Define the best log-concave approximation of $Q_0$ as
\[
	f_0^* = \argmax_{f \in \mathcal{F}} \, \int \log f \, dQ_0\,,
\]
where $\mathcal{F}$ is the class of log-concave densities. If $Q_0$ has a log-concave density function $f_0$, then $f_0^* = f_0$. Otherwise, in the case that $f_0$ has finite entropy, $f_0^*$ is the density function that minimizes the Kullback--Leibler divergence $D_{KL}(f_0,f) = \int f_0 \log(f_0/f)$ over all $f \in \mathcal{F}$. Consequently, if $f_0$ is not too far away from log-concave, $f_0^*$ will be reasonably close to $f_0$. More details regarding the properties of $f_0^*$ can be found in \citet{CuleSamworth2010}, \citet{DSS2011} and \citet{ChenSamworth2013}.

Now we are in the position to state the consistency theorem.
\begin{thm}[Consistency in ARMA]
\label{Thm:armaconsistency}
Let $(\hat{f}_n, \hat{\boldsymbol{\theta}}_n)$ be a sequence of LCMLEs defined in (\ref{Eq:armamle2}). Under assumptions \textbf{(A.1)}--\textbf{(A.5)}, almost surely
\begin{align}
\label{Eq:armapar}
	\int \bigl| \hat{f}_n(x) - f_0^*(x) \bigr| \, dx  \rightarrow 0
	\quad \mbox{ and } \quad
	\hat{\boldsymbol{\theta}}_n \rightarrow  \boldsymbol{\theta}_0, 
\end{align}
as $n \rightarrow \infty$.
\end{thm}

\noindent\textbf{Remarks}: 
\begin{enumerate}[1.]
\setlength{\itemsep}{0pt}
\setlength{\parskip}{0pt}
\setlength{\parsep}{0pt}
\item It is \emph{possible} to drop the first part of condition \textbf{(A.1)} (i.e. $Q_0$ has a density function), and replace it by the following slightly weaker condition:
\begin{enumerate}
\setlength{\itemsep}{0pt}
\setlength{\parskip}{0pt}
\setlength{\parsep}{0pt}
\item[\textbf{(A.1*)}] $Q_0$ is non-degenerate and has finite first moment. 
\end{enumerate}
But then the density part of the LCMLE exists only with asymptotic probability one. See also the numerical experiments in Section~\ref{Sec:simae} for more evidence. 

\item The convergence of $\hat{f}_n(x)$ in the $L_1$ norm can be strengthened as follows: suppose that $a:\mathbb{R} \rightarrow \mathbb{R}$ is a sublinear function, i.e. $a(x+y) \le a(x) + a(y)$ and $a(rx) = ra(x)$ for all $x,y \in \mathbb{R}$ and $r \ge 0$, satisfying $e^{a(x)}f_0^*(x) \rightarrow 0$ as $|x| \rightarrow \infty$. Then it can be shown that under the conditions of Theorem~\ref{Thm:armaconsistency}, 
\[
	\int e^{a(x)}|\hat{f}_n(x) - f_0^*(x)| \rightarrow 0, \quad a.s.
\]
\citep[Theorem~2.1]{SHD2011}. 

\item Unlike the common approaches in the literature, we do not require the variance of $Q_0$ to be finite in order to establish the consistency of $\hat{\boldsymbol{\theta}}_n$ for the LCMLE. For other estimator that can handle the infinite variance ARMA, see \citet{PWY2007}.
\end{enumerate}

Theorem~\ref{Thm:armaconsistency} states that the parametric part of the LCMLE is consistent even if $Q_0$ is not log-concave. This is somewhat surprising because one would have thought that imposing incorrect shape constraints would lead to asymptotic biases in estimating $\boldsymbol{\theta}_0$.  We stress that techniques developed in \citet{DSS2011}, especially their Theorem~3.5, play important roles in this proof. To help the reader better understand the result, here we briefly outline its main ideas in the simplest AR(1) setting: 
\begin{enumerate}[1.]
\setlength{\itemsep}{0pt}
\setlength{\parskip}{0pt}
\setlength{\parsep}{0pt}
\item The initial value $X_0$ is asymptotically unimportant.
\item By the empirical process theory for stationary and ergodic sequences, it can be shown that 
\[ 
	\sup_{a_1 \in \Theta} \left| \sup_{\phi \in \Phi_0} l_n(\phi,a_1) - \sup_{\phi \in \Phi_0} \mathbb{E}\phi(X_2 - a_1 X_1) \right| \stackrel{a.s.}{\rightarrow} 0, \quad \mbox{ as }  n \rightarrow \infty.
\]
\item Because of the structure of AR(1), we can rewrite $X_2 - a_1 X_1$ as $\epsilon_2 + (a_{01}-a_1) X_1$. Since $\epsilon_2$ and $X_1$ are independent, one may appeal to Theorem~3.5 of \citet{DSS2011} to see that at the ``distributional'' level, $\sup_{\phi \in \Phi_0} \mathbb{E}\phi(X_2 - a_1 X_1)$ achieves its \emph{unique} maximum at $a_1 = a_{01}$. Note that we do \emph{not} require the distribution of $\epsilon_2$ or $X_1$ to be log-concave in order to enforce Theorem~3.5 of \citet{DSS2011}.
\item As $\Theta$ is compact, the consistency of the parametric part can be established using a standard compactness argument. We emphasize that the consistency does not rely on the correct specification of the shape restrictions (which is inherited from the previous point).
\end{enumerate}

When $q=0$, there is no need to estimate the innovations iteratively, so assumptions can be relaxed to derive a consistent LCMLE. 
\begin{cor}
\label{Cor:arconsistency}
Let $(\hat{f}_n, \hat{\boldsymbol{\theta}}_n)$ be a sequence of LCMLEs defined in (\ref{Eq:armamle2}). If $q=0$, then under assumptions \textbf{(A.1)}, \textbf{(A.2)} and \textbf{(A.4)}, (\ref{Eq:armapar}) holds almost surely.
\end{cor}

\subsection{The smoothed log-concave maximum likelihood estimator}
\label{Sec:linearsmooth}
In this subsection, we describe a variant of the LCMLE. It has some superior properties over the LCMLE defined in (\ref{Eq:armamle2}), is easy to implement, and yet remains computationally feasible.

One problem associated with the LCMLE is that the estimated density function $\hat{f}_n$ is not everywhere differentiable on the real line. It is not even continuous on the boundary of its support. In fact, non-smoothness is a characteristic feature of shape-constrained maximum likelihood estimators. 

To build an estimator with more attractive visual appearance, and to offer potential improvement in small sample sizes, \citet{DumbgenRufibach2009} introduced a smoothed (yet still fully automatic) version of the univariate log-concave maximum likelihood density estimator via convolving with a Gaussian density. \citet{ChenSamworth2013} extended this idea to the multivariate setting and studied its theoretical properties.

In the case that $Q_0$ has finite variance, we can adapt this general idea by modifying Step (ii) of the ARMA estimation procedure as follows:
\begin{enumerate}
\setlength{\itemsep}{0pt}
\setlength{\parskip}{0pt}
\setlength{\parsep}{0pt}
\item[(ii)] Define the empirical innovation distribution
\[
	\tilde{Q}_{n,\hat{\boldsymbol{\theta}}_n} = \frac{1}{n} \, \sum_{t=1}^n \delta_{\tilde{\epsilon}_t(\hat{\boldsymbol{\theta}}_n)} \ ,
\]
where $\delta_a$ denotes a Dirac point mass at $a$. Let $\tilde{f_n} = \hat{f_n} \star \phi_{\hat{A}_n}$ with 
\[
	\hat{A}_n = \int x^2 d \tilde{Q}_{n,\hat{\boldsymbol{\theta}}_n}(x) - \int x^2 \hat{f}_n(x) dx,
\]
where `$\star$' is the convolution operator and $\phi_{A}$ is the univariate normal density with mean zero and variance $A$. 
Return $\tilde{f_n}$ and the same $\hat{\boldsymbol{\theta}}_n$. We call $(\tilde{f_n}, \hat{\boldsymbol{\theta}}_n)$ the \emph{smoothed log-concave maximum likelihood estimator for ARMA} or simply the \emph{smoothed LCMLE}.
\end{enumerate}

It can be shown that $\hat{A}_n$ is always positive, so $\tilde{f_n}$ is well-defined. We note that the value of $\hat{\boldsymbol{\theta}}_n$ remains unchanged, but now $\hat{f}_n$ is replaced by its slightly smoothed version $\tilde{f}_n$. All the theoretical results described in Section~\ref{Sec:lineartheory} are still valid. But instead of converging to $f_0^*$ in Theorem~\ref{Thm:armaconsistency} and Corollary~\ref{Cor:arconsistency}, $\tilde{f}_n$ converges to $f_0^{**}$, i.e. $\int |\hat{f}_n(x)-f_0^{**}(x)| dx \stackrel{a.s.}{\rightarrow} 0$, where $f_0^{**}= f_0^* \star \phi_{A^*}$ with $A^* = \int x^2 f_0(x) dx - \int x^2 f_0^*(x)dx$ (cf. \citet{ChenSamworth2013}). Nevertheless, in the case that $f_0$ is log-concave, $f_0^{**}=f_0^*=f_0$.

\section{ARMA-GARCH models}
\label{Sec:nonlinear}
The class of ARCH models was developed by \citet{Engle1982} and generalized by \citet{Bollerslev1986}. It is common in practice to fit ARMA models with GARCH errors, which can be viewed as an extension of both ARMA and GARCH models. See \citet{FrancqZakoian2010} for a nice introduction. 

We write the ARMA($p,q$)-GARCH($r,s$) model as
\begin{align*}
	X_t   		&= \sum_{i=1}^p a_i X_{t-i} + \sum_{i=1}^q b_i \eta_{t-i} + \eta_t, \\
	\eta_t 		&= \sigma_t \epsilon_t, \\
	\sigma_t^2 	&= c + \sum_{i=1}^r \alpha_{i} \eta_{t-i}^2 + \sum_{i=1}^s \beta_{i} \sigma_{t-i}^2 \, ,
\end{align*}
where innovations $\{\epsilon_t\}$ are \emph{i.i.d.} random variables with \emph{unit} second moment (i.e. $\mathbb{E}\epsilon^2_t = 1$). Here $c > 0$, $\alpha_i \ge 0$ for $i = 1,\ldots,r$ and $\beta_i \ge 0$ for $i = 1,\ldots,s$.

A primary feature of this class of models is that it allows the conditional variance of the errors to change over time. Often the distribution of $\{\epsilon_t\}$ is assumed to be standard normal, so that estimates of the parameters can be derived by maximizing the conditional log-likelihood. If the distribution of $\{\epsilon_t\}$ is misspecified, maximizing the Gaussian quasi-log-likelihood still gives consistent estimates of these parameters \citep{FrancqZakoian2004}, but is occasionally  inefficient. Non-Gaussian quasi-maximum likelihood estimators also exist in the literature, but they may lead to inconsistent estimates if the distribution of the innovation is misspecified \citep{NeweySteigerwald1997}. In the following, we tackle the problem by assuming that the innovations $\{\epsilon_t\}$ have a log-concave density.

\subsection{The log-concave maximum likelihood estimator}
\label{Sec:nonlinearconstruct}
Suppose that the observations $X_1, \ldots, X_n$ constitute a realization of an ARMA($p,q$)-GARCH($r,s$) process, where the orders $p$, $q$, $r$ and $s$ are assumed to be known. The vector of the parameters 
\[
	\boldsymbol{\theta} = (\mathbi{a}^T,\mathbi{b}^T, c,\boldsymbol{\alpha}^T,\boldsymbol{\beta}^T)^T = (a_1, \ldots, a_p,b_1,\ldots,b_q, c, \alpha_1, \ldots, \alpha_r,\beta_1,\ldots,\beta_s)^T
\]
belongs to a parameter space of form $\Theta \subseteq \mathbb{R}^{p+q} \times (0,\infty) \times [0,\infty)^{r+s}$.

Both the true distribution of $\{\epsilon_t\}$ and the true value of the parameter vector are unknown and to be estimated. They are denoted respectively by $Q_0$ and 
\[
	\boldsymbol{\theta}_0 = (\mathbi{a}_0^T,\mathbi{b}_0^T, c_0,\boldsymbol{\alpha}_0^T,\boldsymbol{\beta}_0^T)^T = (a_{01}, \ldots, a_{0p},b_{01},\ldots,b_{0q}, c_0, \alpha_{01}, \ldots, \alpha_{0r},\beta_{01},\ldots,\beta_{0s})^T.
\] 

In order to construct the LCMLE, we impose the following conditions:
\begin{enumerate}
\setlength{\itemsep}{0pt}
\setlength{\parskip}{0pt}
\setlength{\parsep}{0pt}
\item[\textbf{(B.1)}] $Q_0$ has unit second moment and a density function $f_0$; 
\item[\textbf{(B.2)}] $\boldsymbol{\theta}_0 \in \Theta$ and $\Theta$ is compact;
\end{enumerate}

\noindent\textbf{Remark}: 
Without loss of generality, we can assume in the rest of the paper that \textbf{(B.2)} holds true when the parameter space is of form
\[
	\Theta =  [-1/\delta, 1/\delta]^{p+q} \times [\delta, 1/\delta] \times [0, 1/\delta]^{r+s} \subseteq \mathbb{R}^{p+q+r+s+1}
\]
for some known sufficiently small $\delta \in (0,1)$.

Now the log-concave log-likelihood of ARMA-GARCH can be expressed as 
\begin{align}
%\label{Eq:armagarch-loglike}
	l_n(\phi, \boldsymbol{\theta}) = l_n(\phi, \boldsymbol{\theta}; X_1, \ldots, X_n) = \frac{1}{n}\sum_{t=1}^n \phi \left( \frac{\tilde{\eta}_t(\boldsymbol{\theta})}{\sqrt{\tilde{\sigma}_t^2(\boldsymbol{\theta})}} \right) - \frac{1}{2n} \sum_{t=1}^n \log \left( \tilde{\sigma}_t^2(\boldsymbol{\theta}) \right),
\end{align}
where $\phi \in \Phi_0$, $\boldsymbol{\theta} \in \Theta$, $\{\tilde{\eta}_t(\boldsymbol{\theta})\}$ and $\{\tilde{\sigma}_t^2(\boldsymbol{\theta})\}$ are defined recursively by
\begin{align*}
	\tilde{\eta}_t(\boldsymbol{\theta})	&=  X_t - \sum_{i=1}^p a_i X_{t-i} - \sum_{i=1}^q b_i \tilde{\eta}_{t-i}(\boldsymbol{\theta}), \\
	\tilde{\sigma}_t^2(\boldsymbol{\theta})	&= c + \sum_{i=1}^r \alpha_i \tilde{\eta}_{t-i}^2(\boldsymbol{\theta}) + \sum_{i=1}^s \beta_i \tilde{\sigma}_{t-i}^2(\boldsymbol{\theta}).
\end{align*}
If $r \ge q$, the required initial values are $X_0, \ldots, X_{1-(r-q)-p}, \tilde{\eta}_{q-r}(\boldsymbol{\theta}),\ldots,\tilde{\eta}_{1-r}(\boldsymbol{\theta}), \tilde{\sigma}_{0}^2(\boldsymbol{\theta}), \ldots, \tilde{\sigma}_{1-s}^2(\boldsymbol{\theta})$; otherwise, they are $X_0,\ldots,X_{1-(r-q)-p},\tilde{\eta}_{0}(\boldsymbol{\theta}),\ldots,\tilde{\eta}_{1-q}(\boldsymbol{\theta}), \tilde{\sigma}_{0}^2(\boldsymbol{\theta}), \ldots, \tilde{\sigma}_{1-s}^2(\boldsymbol{\theta})$. As is shown in the appendix, the choice of these unknown initial values is asymptotically irrelevant to our final estimates. To simplify the analysis, we take them to be fixed. 

Let $\Phi_1$ be a subset of $\Phi$ such that 
\[
	\Phi_1 = \left\{\phi \in \Phi : \int e^{\phi(x)} dx = 1, \int x^2 e^{\phi(x)} dx = 1 \right\}.
\]
Naturally, one would attempt to maximize $l_n(\phi, \boldsymbol{\theta})$ over $\Phi_1 \times \Theta$. However, it is hard to enforce all the constraints simultaneously. Therefore we seek to reformulate the optimization problem. 

Our approach is motivated by the following identifiability property of the ARMA-GARCH process: if we replace $(f_0(\cdot),\mathbi{a}_0,\mathbi{b}_0, c_0,\boldsymbol{\alpha}_0,\boldsymbol{\beta}_0)$ by $(\sqrt{k}f_0(\sqrt{k}\cdot), \mathbi{a}_0,\mathbi{b}_0, k c_0, k \boldsymbol{\alpha}_0,\boldsymbol{\beta}_0)$ for any constant $k \in (0, \infty)$, the ARMA-GARCH process remains unchanged. Therefore we can enforce the constant term to be one in Step~(i) of the following procedure and transform it back in Step~(iii):

\begin{enumerate}
\setlength{\itemsep}{0pt}
\setlength{\parskip}{0pt}
\setlength{\parsep}{0pt}
\item[(i)] Define the \emph{transformed} parameter space 
\begin{align*}
	\Theta'= [-1/\delta, 1/\delta]^{p+q} \times \{1\} \times [0, 1/\delta^2]^r \times [0,1/\delta]^s.
\end{align*}
Let $(\hat{\phi}_n', \hat{\mathbi{a}}_n, \hat{\mathbi{b}}_n, 1, \hat{\boldsymbol{\alpha}}_n',\hat{\boldsymbol{\beta}}_n)$ be a maximizer over $(\phi,\boldsymbol{\theta}) \in \Phi \times \Theta'$ of  
\begin{align}
\label{Eq:armagarchmle1}
	\Lambda_n(\phi, \boldsymbol{\theta}) = \Lambda_n(\phi, \boldsymbol{\theta}; X_1, \ldots, X_n) = \frac{1}{n} \sum_{t=1}^n \phi \left( \frac{\tilde{\eta}_t(\boldsymbol{\theta})}{\sqrt{\tilde{\sigma}_t^2(\boldsymbol{\theta})}} \right)  - \frac{1}{2n} \sum_{t=1}^n \log (\tilde{\sigma}_t^2(\boldsymbol{\theta})) - \int e^{\phi(x)}dx + 1.
\end{align}
For convenience, we denote $(\hat{\mathbi{a}}_n^T, \hat{\mathbi{b}}_n^T, 1, (\hat{\boldsymbol{\alpha}}_n')^T,\hat{\boldsymbol{\beta}}_n^T)^T$ by $\hat{\boldsymbol{\theta}}_n'$.

\item[(ii)] Set
\[
	\hat{c}_n = \frac{1}{n} \sum_{t=1}^n \frac{\tilde{\eta}_t^2(\hat{\boldsymbol{\theta}}_n')}{\tilde{\sigma}_t^2(\hat{\boldsymbol{\theta}}_n')}.
\]

\item[(iii)] Return
\begin{align}
\label{Eq:armagarchmle2}
	\hat{f}_n(x) &= \sqrt{\hat{c}_n} e^{\hat{\phi}_n'(\sqrt{\hat{c}_n} x)} \, \mbox{ and } \, 
	\hat{\boldsymbol{\theta}}_n = (\hat{\mathbi{a}}_n^T, \hat{\mathbi{b}}_n^T, \hat{c}_n, \hat{c}_n (\hat{\boldsymbol{\alpha}}_n')^T,\hat{\boldsymbol{\beta}}_n^T)^T,
\end{align}
where $(\hat{f}_n,\hat{\boldsymbol{\theta}}_n)$ is called the LCMLE of $(f_0,\boldsymbol{\theta}_0)$ in ARMA-GARCH. 
\end{enumerate}

\noindent\textbf{Remarks}: 
\begin{enumerate}[1.]
\setlength{\itemsep}{0pt}
\setlength{\parskip}{0pt}
\setlength{\parsep}{0pt}
\item The function $\hat{f}_n$ is always a probability density function. Though it is not guaranteed that $\int x^2 \hat{f}_n(x) dx = 1$, we show in Section~\ref{Sec:nonlineartheory} that this statement is asymptotically true if $f_0$ is log-concave. 
\item By making use of the smoothed log-concave density estimator, it is easy to modify the above steps to enforce the second moment of the estimated innovation distribution to be \emph{exactly} one. See Section~\ref{Sec:nonlinearsmooth} for more details.
\item By setting $p=q=0$, the above procedure can be used for pure GARCH processes. 
\end{enumerate}

\subsection{Theoretical properties}
\label{Sec:nonlineartheory}
%The existence of the LCMLE under the ARMA-GARCH setting is stated in the next theorem.
\begin{thm}[Existence in ARMA-GARCH]
\label{Thm:armagarchexist}
For every $n > p+q+r+s+1$, under assumptions \textbf{(B.1)} -- \textbf{(B.2)}, the LCMLE $(\hat{f}_n, \hat{\boldsymbol{\theta}}_n)$ defined in (\ref{Eq:armagarchmle2}) exists with probability one.
\end{thm}

In addition to the ARMA polynomials mentioned in Section~\ref{Sec:linear}, we define the GARCH polynomials as
\[
	\mathcal{A}_{\boldsymbol{\theta}}(z) = \sum_{i=1}^r \alpha_i z^i \quad \mbox{ and } \quad \mathcal{B}_{\boldsymbol{\theta}}(z) = 1 - \sum_{i=1}^s \beta_i z^i.
\]
To show strong consistency, several mild assumptions are needed:
\begin{enumerate}
\setlength{\itemsep}{0pt}
\setlength{\parskip}{0pt}
\setlength{\parsep}{0pt}
\item[\textbf{(B.3)}] For all $\boldsymbol{\theta} \in \Theta$, $\sum_{i=1}^s \beta_i < 1$.
\item[\textbf{(B.4)}] The GARCH($r,s$) process with the innovation distribution $Q_0$ and the parameter vector $(c_0, \boldsymbol{\alpha}_0^T,\boldsymbol{\beta}_0^T)^T$ is strictly stationary and ergodic; 
\item[\textbf{(B.5)}] If $s > 0$, $\mathcal{A}_{\boldsymbol{\theta}_0}(z)$ and $\mathcal{B}_{\boldsymbol{\theta}_0}(z)$ have no common roots, $\mathcal{A}_{\boldsymbol{\theta}_0}(1) \ne 0$ and $\alpha_{0r} + \beta_{0s} \ne 0$.
\end{enumerate}

\noindent\textbf{Remarks}: 
\begin{enumerate}[1.]
\setlength{\itemsep}{0pt}
\setlength{\parskip}{0pt}
\setlength{\parsep}{0pt}
\item It can be shown that the assumption \textbf{(B.3)} is weaker than assuming strict stationarity of the GARCH processes over $\Theta$. For instance, see Corollary~2.2 of \citet{FrancqZakoian2010}.
\item A necessary and sufficient condition for the assumption \textbf{(B.4)} was established by \citet{BougerolPicard1992} in terms of the top Lyapunov exponent. A more interpretable sufficient condition was given by \citet{Bollerslev1986}, namely, $\sum_{i=1}^r \alpha_{0i} + \sum_{i=1}^s \beta_{0i} < 1$. Note that Bollerslev's condition excludes IGARCH and implies second-order stationarity of GARCH, but here we do not need such a strong condition to establish the consistency of our LCMLE.
\item Assumption \textbf{(B.5)} ensures that the GARCH part of the model is identifiable. This assumption also allows for an overidentification of either $r$ or $s$. We refer to Remark~2.4 of \citet{FrancqZakoian2004} for a detailed discussion.
\end{enumerate}

\begin{thm}[Consistency in ARMA-GARCH]
\label{Thm:armagarchconsistency}
Let $(\hat{f}_n, \hat{\boldsymbol{\theta}}_n)$ be a sequence of LCMLEs given by (\ref{Eq:armagarchmle2}). Under assumptions \textbf{(B.1)}--\textbf{(B.5)} and \textbf{(A.4)}--\textbf{(A.5)}, almost surely
\[
	\int \bigl| \hat{f}_n(x) - f_0^*(x) \bigr| \, dx  \rightarrow  0
	\quad \mbox{and} \quad
	\hat{\boldsymbol{\theta}}_n  \rightarrow \boldsymbol{\theta}_0 ,
\]
as $n \rightarrow \infty$. Moreover, if $f_0$ is log-concave, then
\begin{align}
\label{Eq:armagarchvariance}
	\int x^2 \hat{f}_n(x) dx \rightarrow 1, \quad \mbox{a.s.}
\end{align}
\end{thm}

\noindent\textbf{Remarks}: 
\begin{enumerate}[1.]
\setlength{\itemsep}{0pt}
\setlength{\parskip}{0pt}
\setlength{\parsep}{0pt}
\item In the above theorem, \textbf{(B.1)} can be replaced by the following weaker condition:
\begin{enumerate}
\setlength{\itemsep}{0pt}
\setlength{\parskip}{0pt}
\setlength{\parsep}{0pt}
\item[\textbf{(B.1*)}] $\mathbb{E}\epsilon_t^2 = 1$ and there exists no set $\Omega$ of cardinality less than or equal to 2 such that $P(\epsilon_t \in \Omega) = 1$. 
\end{enumerate}
Under \textbf{(B.1*)}, Theorem~\ref{Thm:armagarchexist} no longer holds. Still, one can show that the LCMLE exists with high probability for sufficiently large $n$.

\item It was shown by \citet{FrancqZakoian2004} that the GQMLE for ARMA-GARCH is inconsistent if $\mathbb{E}\epsilon_t \ne 0$. However, this condition is \emph{not} required here to ensure the consistency of our LCMLE. 
\end{enumerate}

We note that there are some similarities between the proofs of Theorem~\ref{Thm:armaconsistency} and Theorem~\ref{Thm:armagarchconsistency}, mainly due to the ARCH($\infty$) presentation of GARCH. However, there are two distinct differences:
\begin{enumerate}[1.]
\setlength{\itemsep}{0pt}
\setlength{\parskip}{0pt}
\setlength{\parsep}{0pt}
\item Because of the nonlinear nature of ARMA-GARCH, a few new tools, notably, Theorem~\ref{Thm:timesnoisel} and Corollary~\ref{Cor:addtimesnoisel}, have been developed to exploit the properties of the log-concave approximation. These results deepen our understanding of this topic and can be found in the appendix. 
\item Here one also needs to handle the extra logarithmic term in (\ref{Eq:armagarchmle1}).
\end{enumerate}

\subsection{The smoothed log-concave maximum likelihood estimator}
\label{Sec:nonlinearsmooth}
Analogous to Section~\ref{Sec:linearsmooth}, the idea of smoothing can be adapted to Step~(iii) of the ARMA-GARCH estimation procedure by changing it as follows:
\begin{enumerate}
\setlength{\itemsep}{0pt}
\setlength{\parskip}{0pt}
\setlength{\parsep}{0pt}
\item[(iii)] Compute $(\hat{f_n}, \hat{\boldsymbol{\theta}}_n)$ in the same way as before.  Set $\hat{A}_n = 1 - \int x^2 \hat{f}_n(x)dx$ and $\tilde{f}_n = \hat{f_n} \star \phi_{\hat{A}_n}$ (N.B. one can prove $\hat{A}_n > 0$). Return $\tilde{f}_n$ and the same $\hat{\boldsymbol{\theta}}_n$. We call $(\tilde{f_n}, \hat{\boldsymbol{\theta}}_n)$ the smoothed LCMLE for ARMA-GARCH.
\end{enumerate}

One nice feature of this new estimator is that the unit second moment constraint is always satisfied, i.e. $\int x^2 \tilde{f}_n(x) dx \equiv 1$. Again, Theorem~\ref{Thm:armagarchexist} and Theorem~\ref{Thm:armagarchconsistency} are still valid, but $\tilde{f}_n$ converges to $f_0^{**}$ instead of $f_0^{*}$ in Theorem~\ref{Thm:armagarchconsistency}.

\section{Computational issues and numerical properties}
\subsection{Computational issues}
\label{Sec:computing}
Computing the LCMLEs proposed in Section~\ref{Sec:linear} and Section~\ref{Sec:nonlinear} is fast and straightforward, especially when the orders of the processes are not too high. To see this, we note that the parametric part of the LCMLEs can be expressed as 
\[
	\hat{\boldsymbol{\theta}}_n \in \argmax_{\theta \in \Theta} \Upsilon_n(\boldsymbol{\theta}) \quad \mbox{ or } \quad
	\hat{\boldsymbol{\theta}}_n \in \argmax_{\theta \in \Theta'} \Upsilon_n(\boldsymbol{\theta}) 
\]
with $\Upsilon_n(\boldsymbol{\theta}) = \sup_{\phi \in \Phi} \Lambda_n(\phi,\boldsymbol{\theta})$. It is shown in the appendix that $\Upsilon_n(\boldsymbol{\theta})$ is a continuous function. Therefore, the optimization problem can be divided into two parts:
\begin{enumerate}
\setlength{\itemsep}{0pt}
\setlength{\parskip}{0pt}
\setlength{\parsep}{0pt}
\item  for a given \emph{fixed} $\boldsymbol{\theta}$, find $\phi \in \Phi$ that maximizes $\Lambda_n(\phi,\boldsymbol{\theta})$;
\item  for a given continuous function $\Upsilon_n(\boldsymbol{\theta})$ on a finite-dimensional compact set (i.e. $\Theta$ or $\Theta'$), find its maximizer.
\end{enumerate}

The first part can be transformed into a convex optimization problem, where the unique optimum $\phi \in \Phi$ can be found very quickly by an active set algorithm implemented in the R package {\tt logcondens} \citep{DumbgenRufibach2011}. More details on its implementation can be found in \citet{DHR2011}. 

The second part is a continuous function optimization problem. Many well-known optimization algorithms can be utilized, including the downhill simplex algorithm \citep{NelderMead1965}, stochastic search \citep{DSS2013}, and differential evolution \citep{PSL2005}. When initial guesses are needed for $\boldsymbol{\theta}$, one reasonable choice would be the GQMLE of $\boldsymbol{\theta}_0$. 

In the following studies, we used the downhill simplex algorithm for optimization, because it suffices for our purpose and is typically much faster than stochastic search or differential evolution. 

\subsection{Simulation I: varying the types of the processes}
\label{Sec:simgq}
To examine the finite sample performance of our method (in estimating the parametric part of the model), we run simulation experiments on a variety of ARMA, GARCH and ARMA-GARCH models. Both the centered exponential innovations (i.e. $f_0(x) = e^{-x-1}, \, x \ge -1$) and the standard Gaussian innovations (i.e. $f_0(x) = \frac{1}{\sqrt{2\pi}} e^{-x^2/2}, \, x \in \mathbb{R}$) are considered. We set the number of observations $n=1000$. Models that we consider, together with their corresponding true values of parameters are listed in Table~\ref{Tab:modelsim}. These values are picked in such a way that all assumptions listed in Section~\ref{Sec:linear}~and~\ref{Sec:nonlinear} are satisfied.

\begin{table}[ht!]
    \centering
    \begin{tabular}{  l  l  }
    \hline
    Linear models & \\ \hline 
    MA(1): & $b_{01} = 0.5$ \\
    AR(2): & $a_{01} = 0.5, a_{02} = -0.5$ \\
    ARMA(1,1): & $a_{01} = 0.5, b_{01} = 0.5$ \\
    ARMA(3,2): & $a_{01} = 0.75, a_{02} = -0.5, a_{03} = 0.25, b_{01} = 0.75, b_{02} = 0.25$ \\ \hline
    Nonlinear models & \\ \hline
    ARCH(1):   & $c_0 = 2, \alpha_{01} = 0.5$ \\
    ARCH(2):   & $c_0 = 1, \alpha_{01} = 0.5, \alpha_{02} = 0.5$ \\
    GARCH(1,1): & $c_0 = 1, \alpha_{01} = 0.25, \beta_{01} = 0.5$ \\
    IGARCH(1,1): & $c_0 = 2, \alpha_{01} = 0.5, \beta_{01} = 0.5$ \\
    GARCH(3,2): & $c_0 = 0.5, \alpha_{01} = 0.3, \alpha_{02} = 0.1, \alpha_{03} = 0.2, \beta_{01} = 0.2, \beta_{02} = 0.1$ \\
    ARMA(1,1)-IGARCH(1,1): & $a_{01} = 0.5, b_{01} = 0.5, c_0 = 0.5, \alpha_{01} = 0.5, \beta_{01} = 0.5$ \\

    \hline\hline 
    \end{tabular}
\caption{Different time series models considered in the simulation study.}
\label{Tab:modelsim}
\end{table}

The results obtained in 1000 simulations by the LCMLE are given in Table~\ref{Tab:sim} in terms of the estimated root-mean-square error (RMSE). Here RMSE is defined as $\sqrt{\mathbb{E}\|\hat{\boldsymbol{\theta}}_n - \boldsymbol{\theta}_0\|^2_2}$, where $\|\cdot\|_2$ is the Euclidean norm. The estimates from the GQMLE are illustrated for comparison. The \texttt{R} package \texttt{fGarch} \citep{WuertzChalabi2012} is used for computing the GQMLE of the nonlinear models.

\begin{table}[ht!]
    \centering
    \begin{tabular}{ | c | c c | c c | }
    \hline 
    Models & \multicolumn{4}{c|}{Estimated RMSE} \\ \cline{1-5}
           & \multicolumn{2}{c|}{centered exponential} & \multicolumn{2}{c|}{Gaussian} \\ 
                	 & LCMLE & GQMLE & LCMLE & GQMLE \\ \cline{1-5}
    MA(1)      		 & 0.0026 & 0.0282 & 0.0287 & 0.0271 \\
    AR(2)       	 & 0.0034 & 0.0392 & 0.0423 & 0.0395 \\
    ARMA(1,1)   	 & 0.0056 & 0.0497 & 0.0521 & 0.0485 \\
    ARMA(3,2)   	 & 0.1019 & 0.2298 & 0.2519 & 0.2399 \\
    ARCH(1)     	 & 0.1807 & 0.3155 & 0.1686 & 0.1510 \\
    ARCH(2)     	 & 0.1151 & 0.2866 & 0.1656 & 0.1500 \\
    GARCH(1,1)  	 & 0.0972 & 0.4699 & 0.3116 & 0.2754\\
    IGARCH(1,1)  	 & 0.1882 & 0.7686 & 0.4727 & 0.4423 \\
    GARCH(2,3)  	 & 0.1044 & 0.3446 & 0.2254 & 0.2217 \\
    ARMA(1,1)-IGARCH(1,1)& 0.0700 & 0.2588 & 0.1599 & 0.1478 \\
    \hline\hline 
    \end{tabular}
\caption{Estimated root-mean-squared error (RMSE) of the LCMLE and the GQMLE in different models with centered exponential or Gaussian innovations.}
\label{Tab:sim}
\end{table}

These results suggest that if the true innovations are non-Gaussian but log-concave, the LCMLE offers substantial improvement over the GQMLE. Strikingly, the reduction in RMSE varies from 50\% to 90\% in the case where the innovations follow the centered exponential distribution.  Even if the true distribution of the innovations is Gaussian, our LCMLE's performance is still comparable to the GQMLE's, indicating that there is little price one has to pay for only assuming the innovations to be log-concave, rather than Gaussian. 

\subsection{Simulation II: varying the innovation distribution and the sample size}
\label{Sec:simae}
In this subsection, we run a small numerical experiment to study the performance of our LCMLE under different innovation distributions and different sample sizes. We compare our method with the adaptive estimator (AE) proposed by \citet{DKW1997} and the GQMLE in estimating the parametric part of the model. For simplicity, we consider the AR(1) model with the true parameter $a_{01} = 0.5$. Different types of innovations together with their features are listed in Table~\ref{Tab:innovationsim}: 
\begin{table}[ht!]
    \centering
    \small
    \begin{tabular}{  l  c c c c}
    \hline
    Type of the innovations & \multicolumn{3}{c}{Features} \\ \cline{2-4}
         & log-concave & symmetric & \specialcell{discrete\\component}\\\hline 
    (a) Centered log-normal $\log N(0,1) - e^{1/2}$					& \xmark & \xmark & \xmark\\
    (b) Student's $t_3$ 								& \xmark & \cmark & \xmark\\ 
    (c) Mixture of Gaussian \& a point mass $\frac{1}{2}N(0,1)+\frac{1}{2}\delta_0$ 	& \xmark & \cmark & \cmark\\ 
    (d) Centered Binomial $B(2,0.4) - 0.8$						& \xmark & \xmark & \cmark\\ 
    (e) Centered exponential 								& \cmark & \xmark & \xmark\\
    (f) Laplace (double exponential)							& \cmark & \cmark & \xmark\\\hline
    \end{tabular}
\caption{Different types of innovations considered and summary of their features.}
\label{Tab:innovationsim}
\end{table}

\begin{figure}[ht!]
 \centering
 $\begin{array}{c c}
  \includegraphics[scale=0.45]{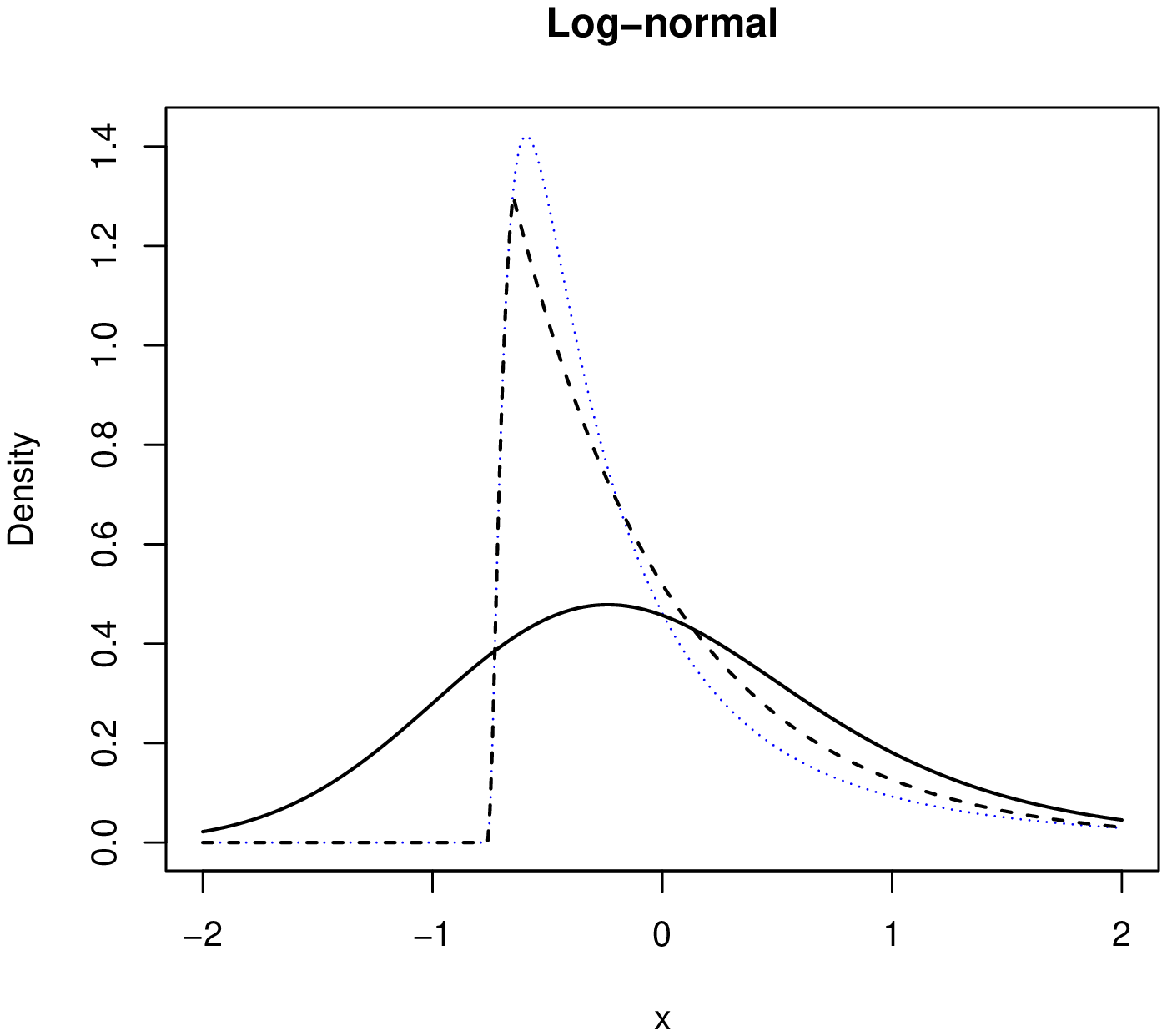} &
  \includegraphics[scale=0.45]{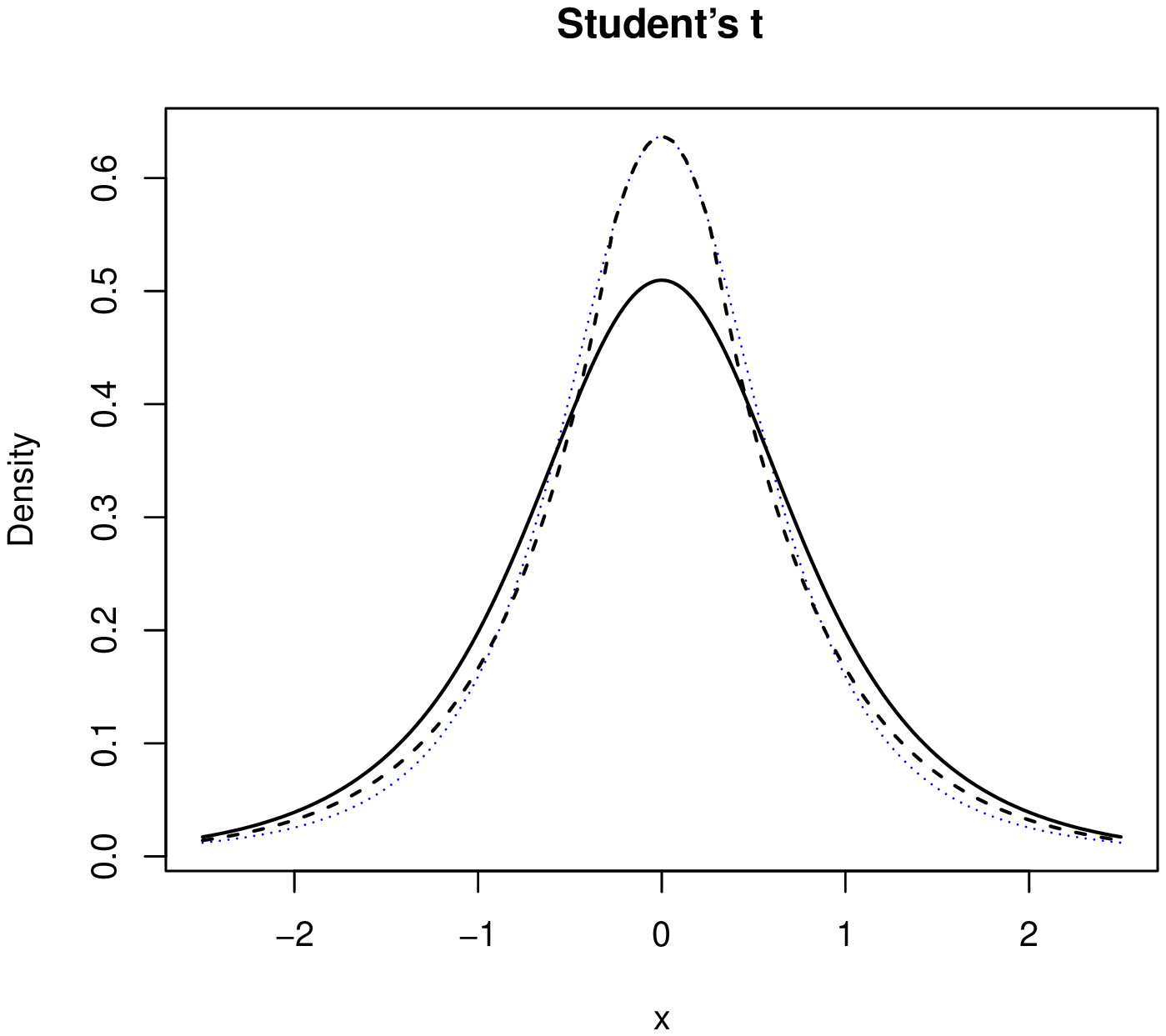} \\
  \mathrm{(a)} & \mathrm{(b)}             \\
  \includegraphics[scale=0.45]{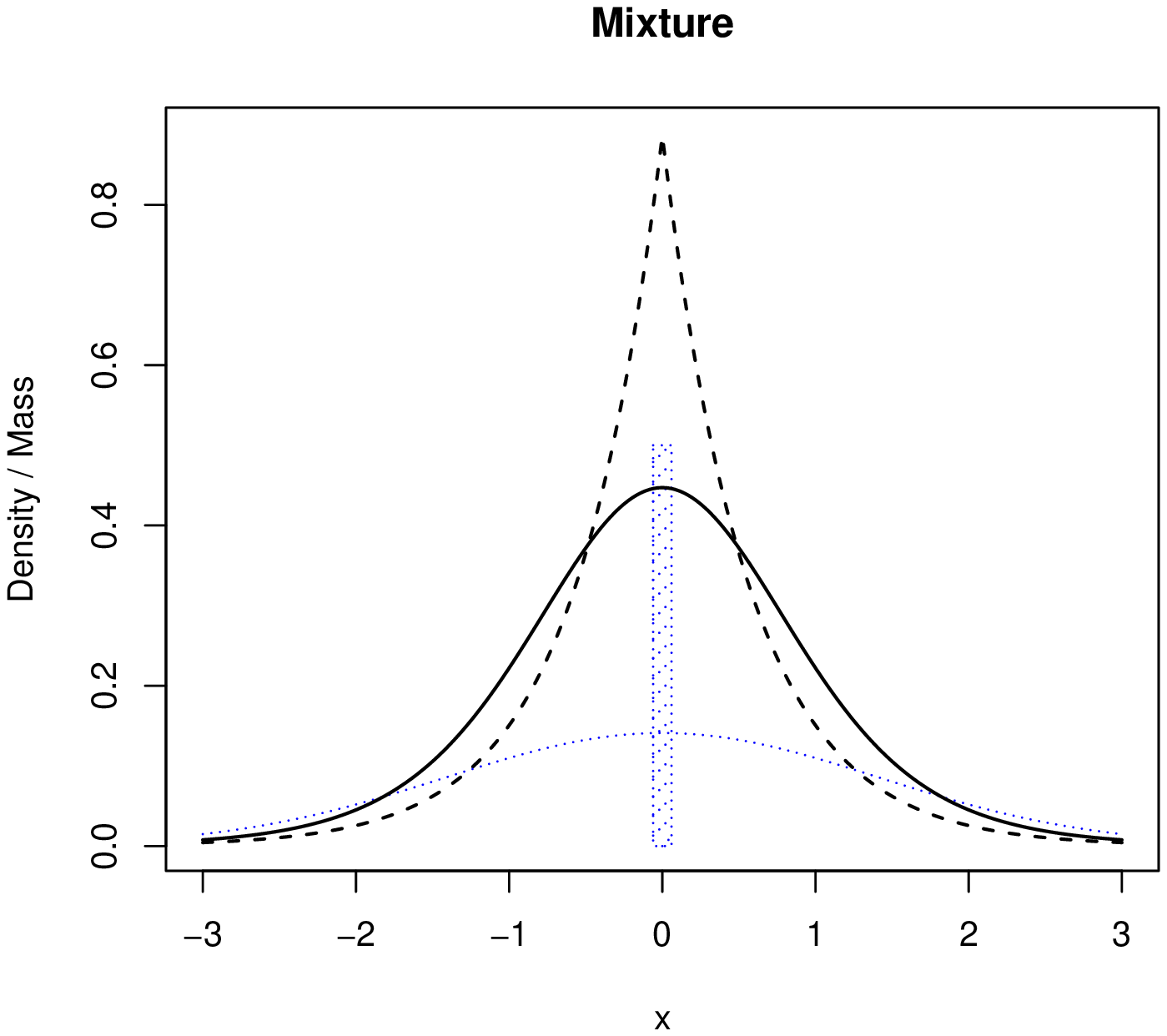} &
  \includegraphics[scale=0.45]{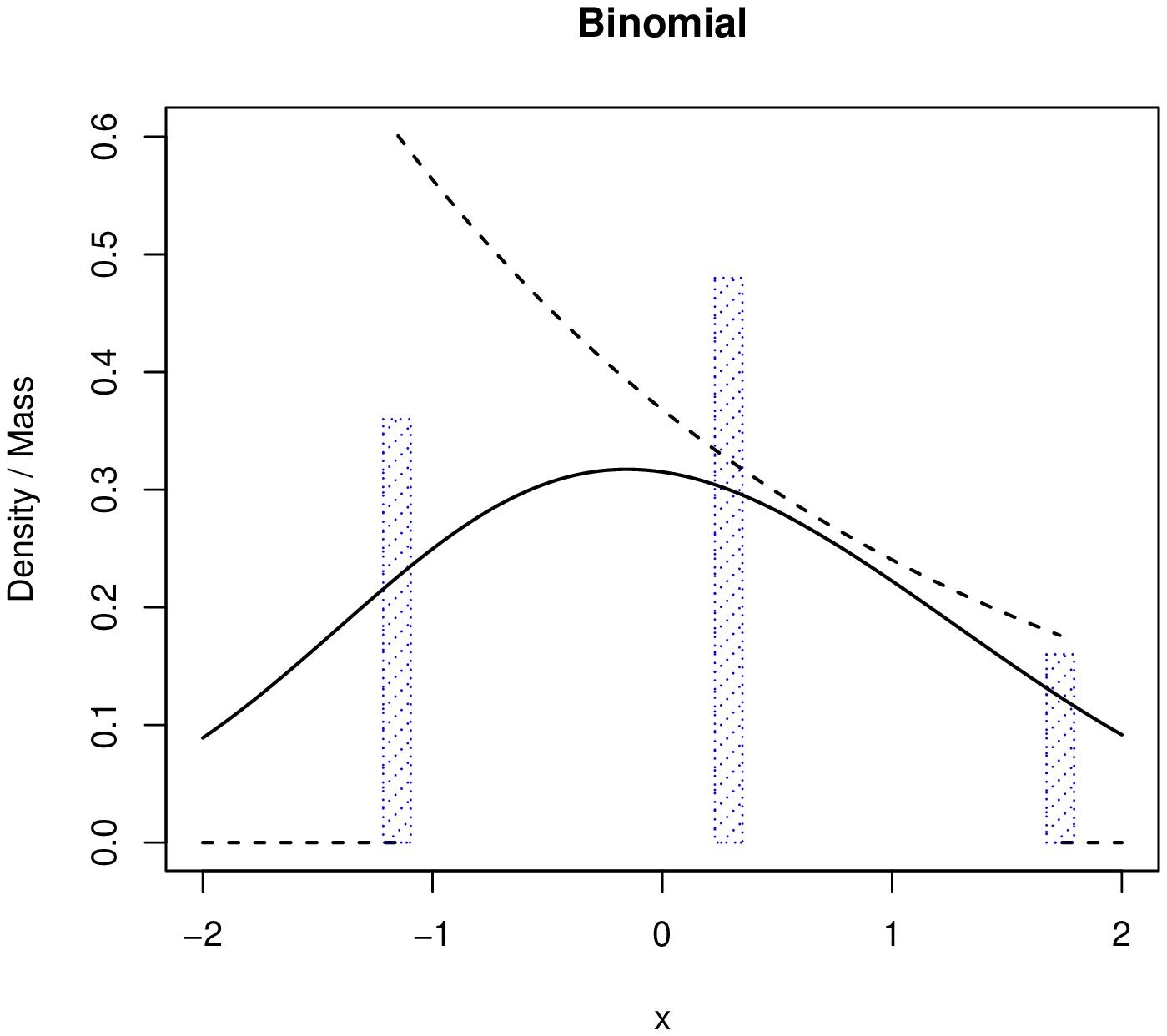} \\
  \mathrm{(c)} & \mathrm{(d)}
 \end{array}$
 \caption{The best log-concave approximation $f_0^*$ and its smoothed analogue $f_0^{**}$ of (a) Centered log-normal; (b) Student's $t_3$;  (c) Mixture of Gaussian \& a point mass;  (d) Centered Binomial. Here $f_0^*$ is plotted in dashed curves, while $f_0^{**}$ is plotted in solid curves. The density function/probability mass function of the innovation distribution $Q_0$ is illustrated in dotted curves or columns. In (c), $Q_0$ consists of continuous and discrete component, these parts are represented respectively by dotted curves and a column. We note that $f_0^*$ is Laplace in (c), and $\log f_0^*$ is linear on $\big[-2\sqrt{3}/3,\sqrt{3}\big]$ in (d).}
 \label{Fig:pseudotruth}
\end{figure}

The innovation distributions in (a)--(d) are not log-concave. Figure~\ref{Fig:pseudotruth} provides information on their corresponding best log-concave approximation $f_0^*$ and the smoothed analogue $f_0^{**}$. For the sake of comparison, we scale the variance of $Q_0$ to one in all scenarios. 

We consider different sample sizes $n=50$, $n=100$ and $n=200$. Small sample sizes are chosen here because the parameter space $\Theta$ is just one-dimensional. Moreover,  no qualitative differences can be observed even if we increase the number of observations to $n = 1000$.

To implement the AE, we use the GQMLE as an initial estimator, together with the kernel density estimator with the Gaussian kernel. Choosing the bandwidth is a tricky task. Although there are theoretical results on the optimal choice of the bandwidth (e.g. see \citet{MammenPark1997} as a starting point), none of them gives practical guidelines on how it would be picked in practice. To address this issue in our simulation study, we use the bandwidth that minimizes the estimated RMSE in each individual situation. This is achieved by considering possible values of the bandwidth on a fine grid and picking the one that minimizes the estimated RMSE. Note that this optimal choice of bandwidth would have been \emph{unknown} in practice. 

The results obtained in 1000 simulations are given in Table~\ref{Tab:adaptive} in terms of the estimated RMSE. Surprisingly, the LCMLE performs substantially better than both the AE and the GQMLE when the innovations have a log-concave but non-Gaussian density. This is quite remarkable because the AE is efficient in the asymptotic sense. We believe this reflects the limitation of the kernel-based methods at small to moderate sample sizes. It is also interesting to witness the robustness of the LCMLE to the misspecification of log-concavity, as the LCMLE outperforms both the AE and the GQMLE in (a) (log-normal) when $n = 50, 100, 200$, and in (b) ($t_3$) when $n = 100, 200$. The most striking improvement of the LCMLE over its competitors occurs in (c) and (d) when the innovation distribution $Q_0$ has discrete component. This is because the adaptation of the AE requires the existence of a density, which is not fulfilled in these cases. Consequently, even though the bandwidth is picked in an optimal manner, the AE can still perform much worse than the LCMLE. Although the asymptotic distributional theory of the LCMLE remains to be investigated, our simulation results have already demonstrated the effectiveness and flexibility of the LCMLE. Finally, we remark that the performance of the GQMLE only depends on the variance of $Q_0$ (in the asymptotic sense, see Chapter~7~and~8 of \citet{BrockwellDavis1991}). The GQMLE's efficiency loss can be quite significant if $Q_0$ is far away from Gaussian. 

\begin{figure}[ht!]
  \centering
  \includegraphics[scale=0.68]{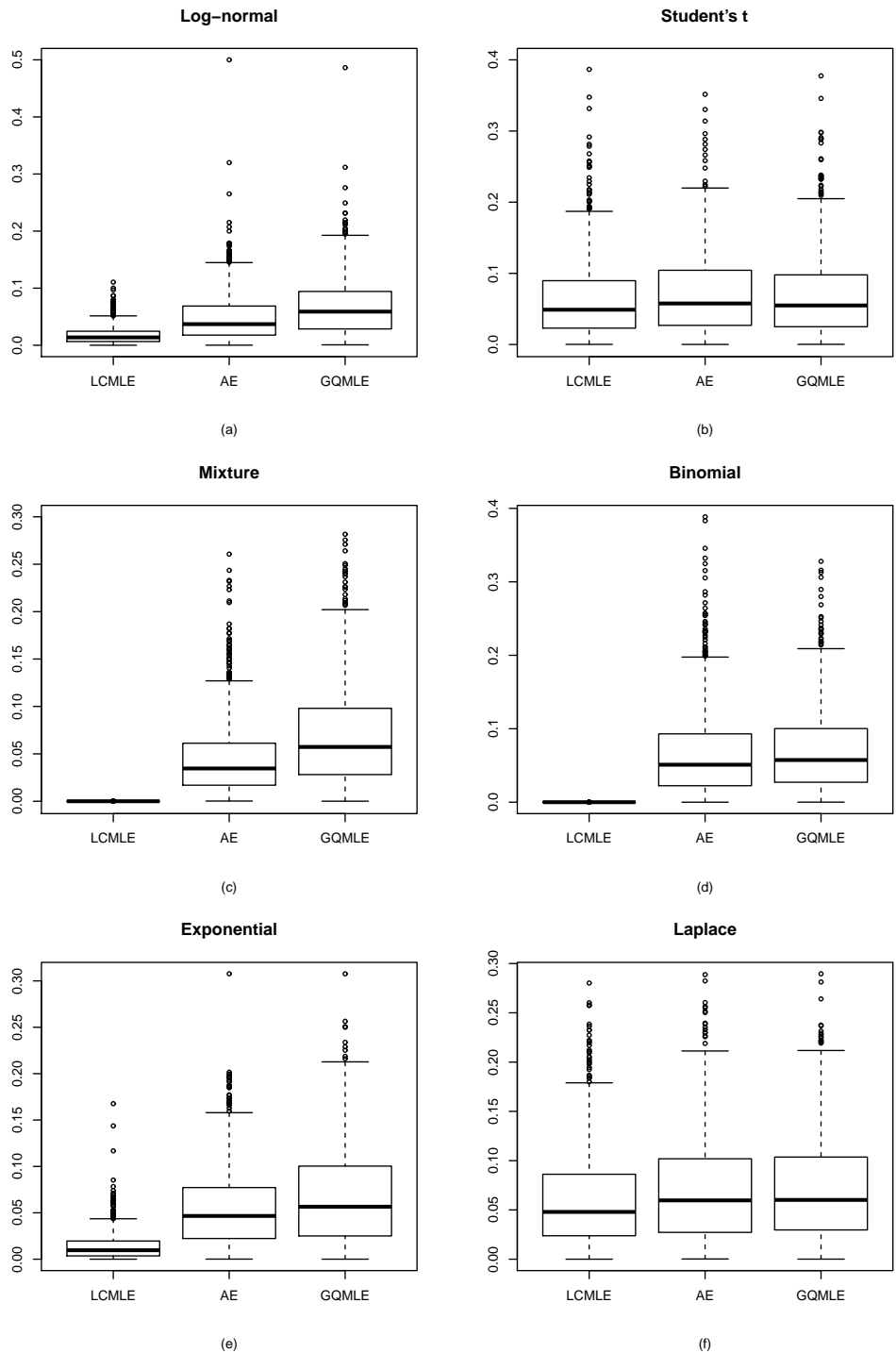}
  \caption{Box plots of the absolute errors for different estimators of $a_{01}$ based on $n = 100$ observations in the setting of AR(1) ($a_{01}=0.5$) with different types of innovations: (a) log-normal; (b) student's $t_3$; (c) mixture of Gaussian and a point mass; (d) centered binomial; (e) centered exponential; (f) Laplace.}
  \label{Fig:arsim1}
\end{figure}

These conclusions are reconfirmed in Figure~\ref{Fig:arsim1}, where box plots of the absolute errors for different estimators of $a_{01}$ based on $n=100$ observations in the above settings are given. Similar conclusions can be obtained under the setting of other ARMA/GARCH/ARMA-GARCH models with different sample sizes.

\begin{table}[ht!]
    \centering
    \begin{tabular}{| c | cccccc | }
    \hline 
    \multicolumn{7}{|c|}{$n=50$} \\ \hline
    $Q_0$: & (a) & (b) & (c) & (d) & (e) & (f) \\ \hline
    LCMLE  & \textbf{0.0417} & 0.1325 & \textbf{0.0237} & $\mathbf{1.5\times 10^{-5}}$ & \textbf{0.0456} & 0.1366  \\
    AE     & 0.1031 & 0.1275 & 0.1026 & 0.1609 & 0.1060 & 0.1243  \\
    GQMLE  & 0.1219 & \textbf{0.1256} & 0.1232 & 0.1266 & 0.1200 & \textbf{0.1228}  \\
    \hline\hline    
    \multicolumn{7}{|c|}{$n=100$} \\ \hline
    $Q_0$: & (a) & (b) & (c) & (d) & (e) & (f) \\ \hline
    LCMLE  & \textbf{0.0240} & \textbf{0.0838} & $\mathbf{1.6 \times 10^{-5}}$ & $\mathbf{1.5\times 10^{-5}}$ & \textbf{0.0212} &  \textbf{0.0793} \\
    AE     & 0.0640 & 0.0899 & 0.0600 & 0.0901 & 0.0694 &  0.0880 \\
    GQMLE  & 0.0839 & 0.0880 & 0.0868 & 0.0884 & 0.0850 &  0.0884 \\
    \hline\hline
    \multicolumn{7}{|c|}{$n=200$} \\ \hline
    $Q_0$: & (a) & (b) & (c) & (d) & (e) & (f) \\ \hline
    LCMLE  & \textbf{0.0144} & \textbf{0.0509} & $\mathbf{1.5 \times 10^{-5}}$ & $\mathbf{1.4 \times 10^{-5}}$ & \textbf{0.0101} & \textbf{0.0530}  \\
    AE     & 0.0422 & 0.0573 & 0.0361 & 0.0513 & 0.0441 & 0.0614  \\
    GQMLE  & 0.0591 & 0.0615 & 0.0625 & 0.0613 & 0.0600 & 0.0658  \\
    \hline\hline
    \end{tabular}
\caption{The estimated RMSE of the LCMLE, the AE (with the optimal choice of bandwidth) and the GQMLE in AR(1) with $n = 50,100,200$ observations. The smallest value in each scenario is highlighted in \textbf{bold}.}
\label{Tab:adaptive}
\end{table}

\subsection{Real data examples}
\label{Sec:realdata}

\subsubsection{Daily log-return of the FTSE 100 index}
\label{Sec:FTSE}
We apply our methodology to the daily log-return of the FTSE 100 index from January 5, 2010 to December 31, 2012 ($n=755$). The GARCH(1,1) model is chosen here because it is by far the most commonly-used model by practitioners. There are also empirical evidences that show the adequacy of modeling the FTSE data by GARCH(1,1). See, for instance, Chapter~8.5 of \citet{FrancqZakoian2010}.   

In order to compare our method with the AE \citep{DrostKlaassen1997}, the following slightly different parameterization of GARCH(1,1) has been used:  
\[
	X_t = \sqrt{c} \epsilon_t \sigma_t, \quad \sigma_t^2 = 1 + \alpha_1' X_{t-1}^2 + \beta_1 \sigma_{t-1}^2, 
\]
where $\{\epsilon_t\}$ are \emph{i.i.d} innovations from a distribution $Q$ with unit second moment. \citet{DrostKlaassen1997} showed that it is possible to adaptively estimate both $\alpha_1'$ and $\beta_1$ under this parameterization. To facilitate the interpretation of the autoregressive parameter $\alpha_1'$, we have standardized the series such that the GQMLE of $c$ equals one. Some key features of the standardized series are summarized in Table~\ref{Tab:FTSEdata}.

\begin{table}[ht!]
    \centering
    \begin{tabular}{ c c c c }
    \hline
    Mean & Standard Deviation & Skewness & Excess Kurtosis \\\hline
    0.0458 & 5.5568 & -0.1404 & 1.8009 \\\hline\hline 
    \end{tabular}
\caption{Estimated characteristics of the standardized series of the FTSE 100 index daily log-return.}
\label{Tab:FTSEdata}
\end{table}

To implement the AE, we use the Gaussian kernel and choose the bandwidth by the heuristic approach suggested in \citet{SunStengos2006}. Their idea is to pick the bandwidth that minimizes the mean squared error (MSE) between the estimated score function and $g'/g$ at the residuals, where $g$ is the density of a target distribution. For simplicity, we select the standard Gaussian as the target distribution. Other choices such as Student's t are also possible, but they do not alter our conclusion. 

\begin{figure}[ht!]
 \centering
 $\begin{array}{c c }
  \includegraphics[scale=0.4]{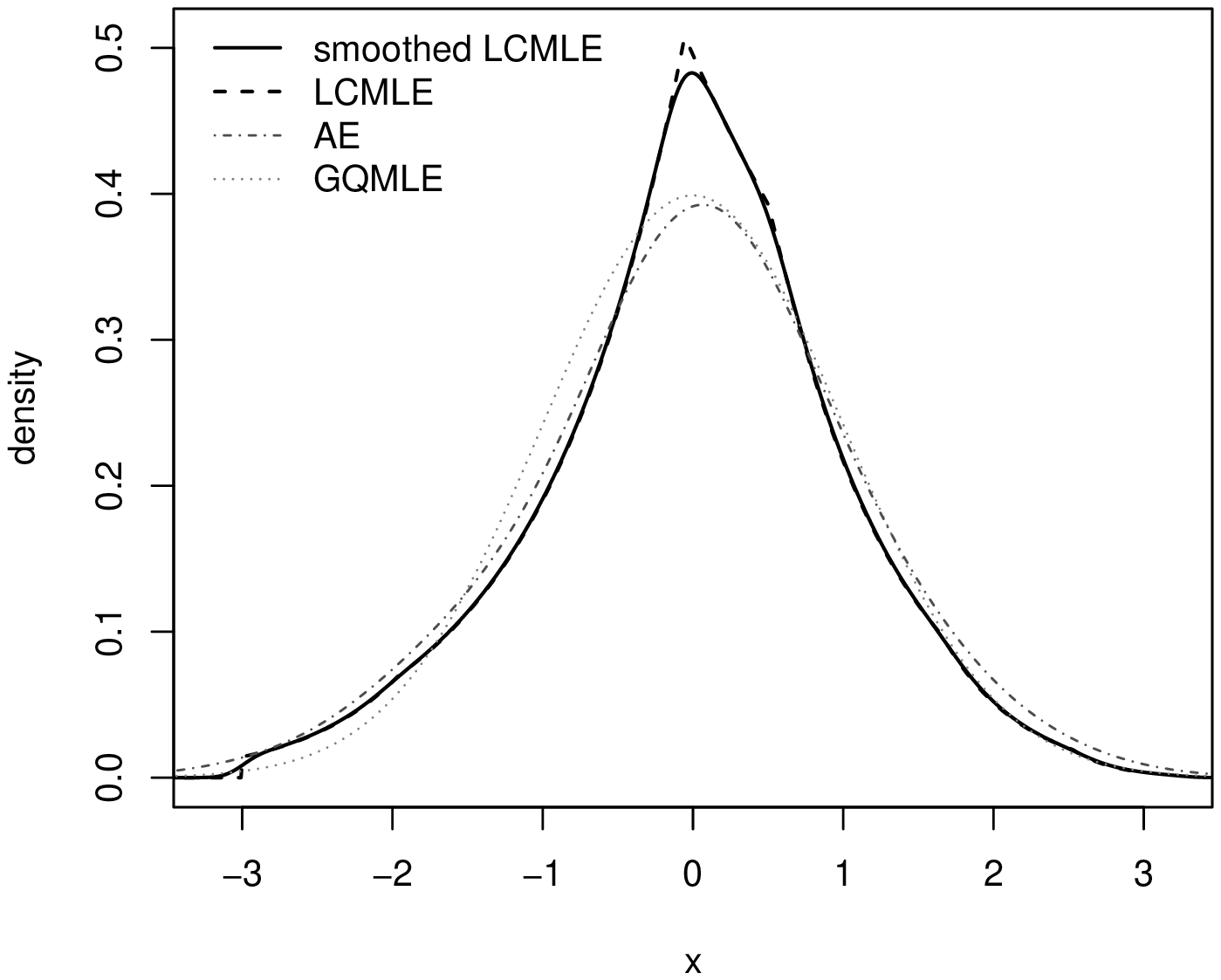} &
  \includegraphics[scale=0.4]{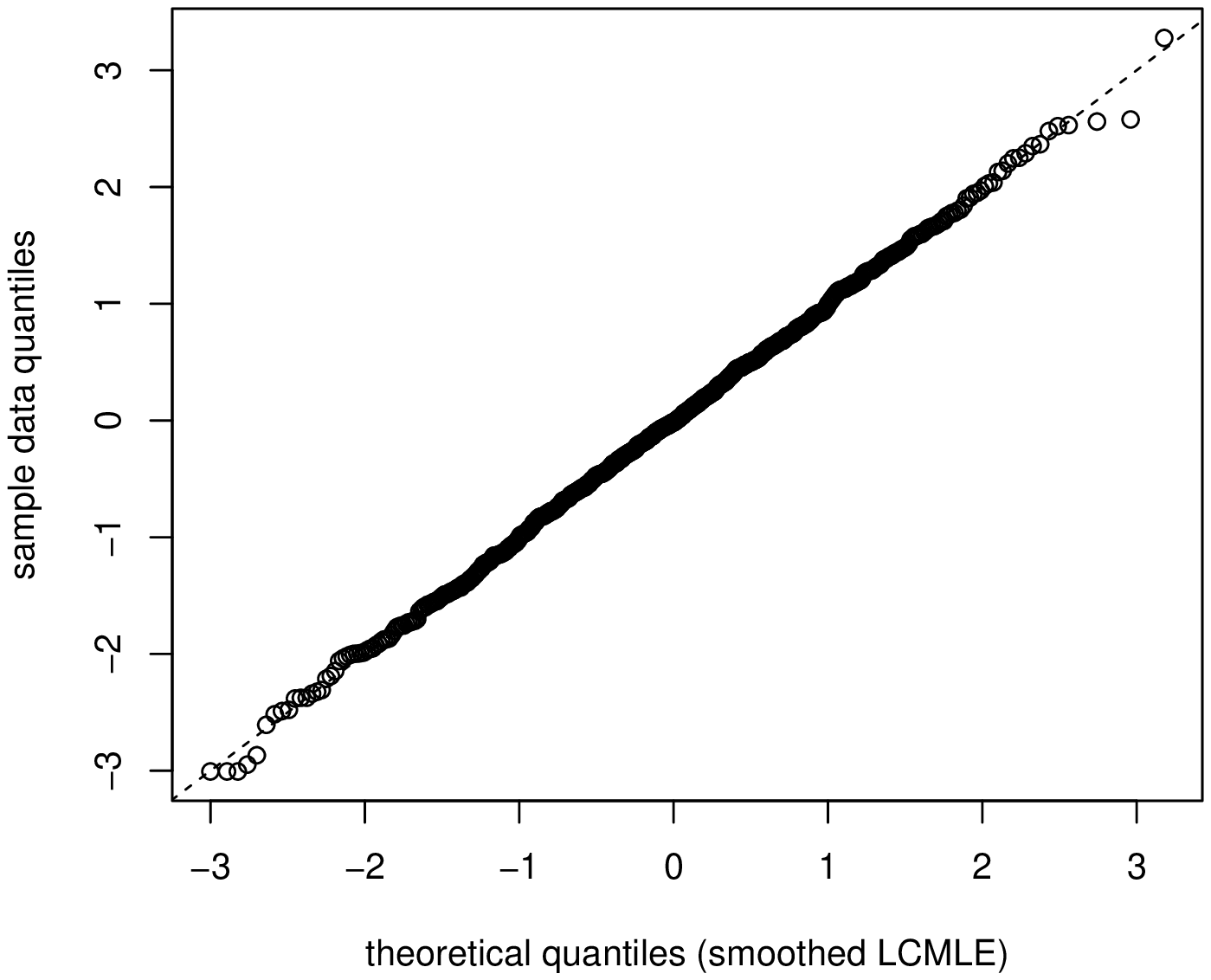}  \\
  \mathrm{(a)} & \mathrm{(b)}  \\
 \end{array}$
 \caption{(a) plots the estimated density functions by the smoothed LCMLE (solid), the LCMLE (dashed), the AE (dash-dotted) and the GQMLE (dotted); (b) gives the Q-Q plot of the residuals against the distribution of the fitted smoothed LCMLE.}
 \label{Fig:FTSE}
\end{figure}

The estimates from the LCMLE, the AE and the GQMLE are given in Table \ref{Tab:FTSEpara}, with the corresponding estimated density functions of $Q$ plotted in Figure~\ref{Fig:FTSE}(a). Among all the fits, the estimated values of the coefficients seem quite similar. In particular, all the methods give estimates of $\beta_1$ greater than 0.8, indicating a strong persistence of shocks on volatility. 

\begin{table}[ht!]
    \centering
    \begin{tabular}{ l c c c }
    \hline
    Method & $\sqrt{c}$ & $\alpha_1'$ & $\beta_1$ \\\hline
    LCMLE: & 0.9663 & 0.1133 & 0.8639  \\
    AE:    & 0.9982 & 0.1692 & 0.8789  \\
    GQMLE: & 1.0000 & 0.1221 & 0.8469  \\\hline\hline 
    \end{tabular}
\caption{Estimated GARCH(1,1) by the LCMLE, the AE and the GQMLE based on the FTSE data.}
\label{Tab:FTSEpara}
\end{table}

However, it can be shown that it is inadequate to modeling this series using Gaussian innovations. In fact, a Shapiro--Wilk test on the residuals gives strong evidence against the normality assumption ($p$-value = 0.006). The estimated density function from the AE visually appears to be close to Gaussian. It is because we have chosen the bandwidth for the purpose of estimating the score function. Often this choice of bandwidth tends to oversmooth the data, so is not necessarily optimal for density estimation. 

On the other hand, our method avoids the issue of choosing the tuning parameters all together. As can be seen from Figure~\ref{Fig:FTSE}(a), the estimated density functions corresponding to both the unsmoothed and smoothed LCMLE demonstrate moderate asymmetric behaviors. Finally, a quantile-quantile (Q-Q) plot of the residuals against the distribution of the fitted smoothed LCMLE is illustrated in Figure~\ref{Fig:FTSE}(b), which implies that the log-concavity assumption on $Q$ is adequate here. 

\subsubsection{Yorkshire rabbit population}
\label{Sec:rabbit}
\begin{figure}[ht!]
 \centering
 $\begin{array}{c c}
  \includegraphics[scale=0.4]{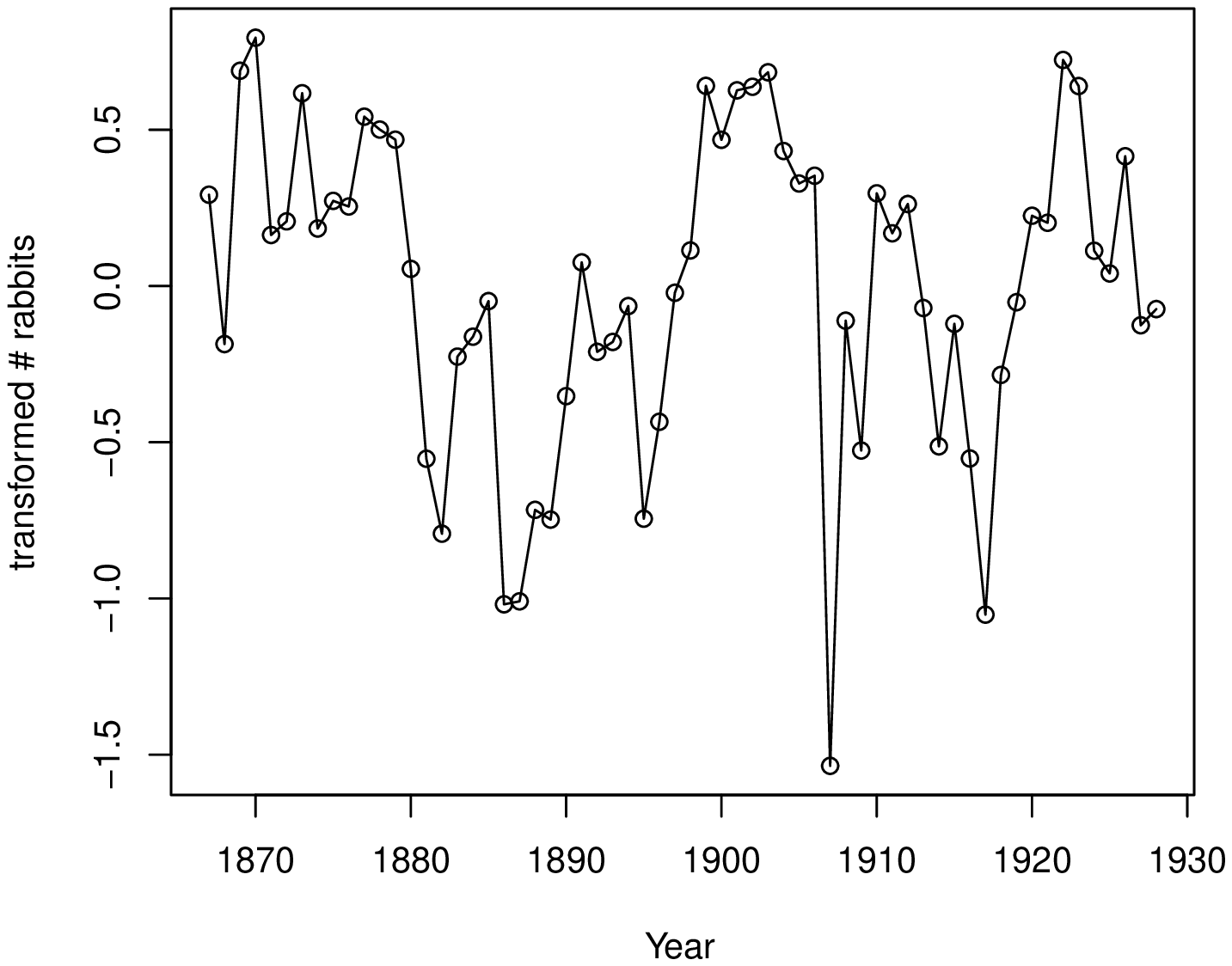} &
  \includegraphics[scale=0.4]{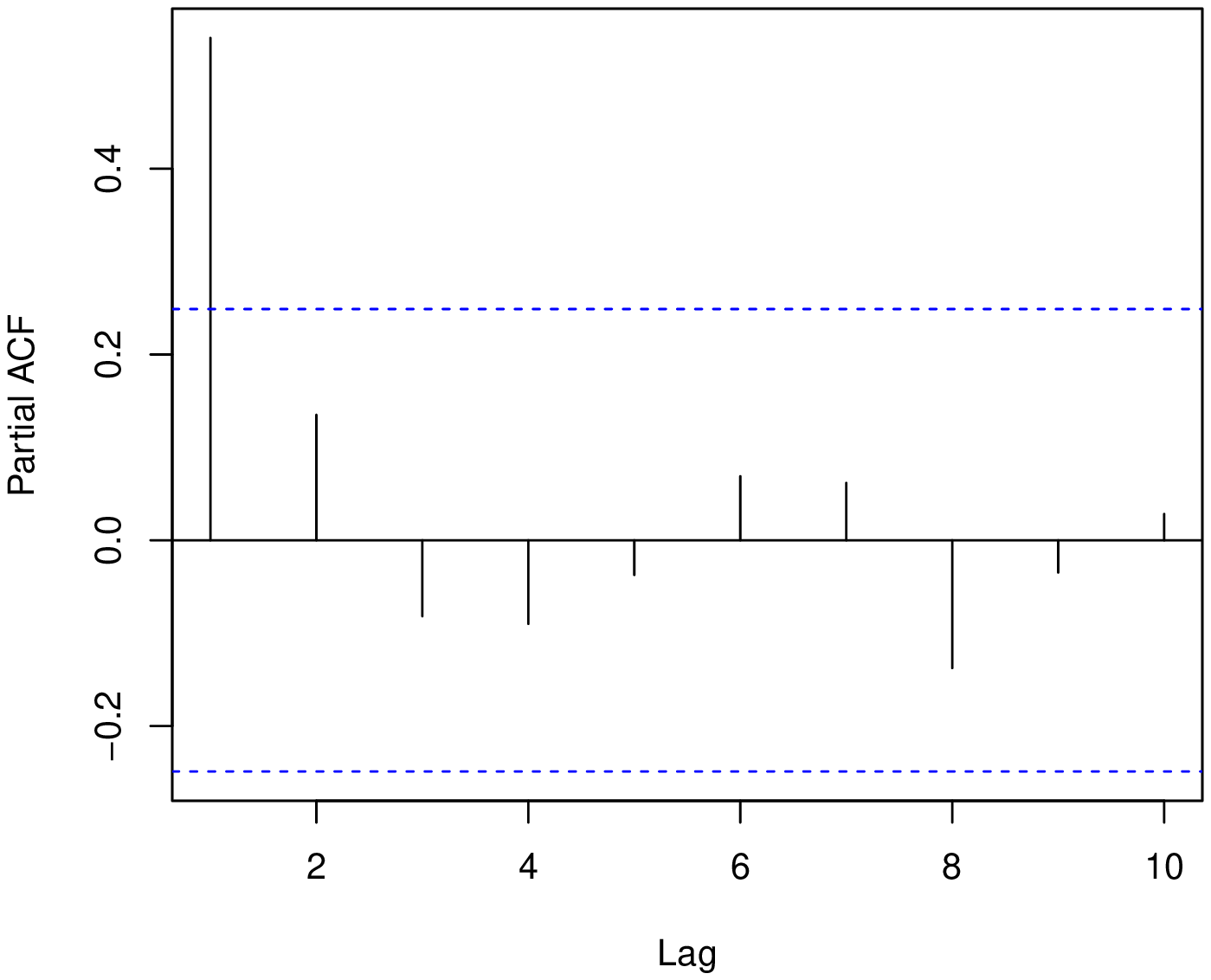} \\
  \mathrm{(a)} & \mathrm{(b)}             \\
  \includegraphics[scale=0.4]{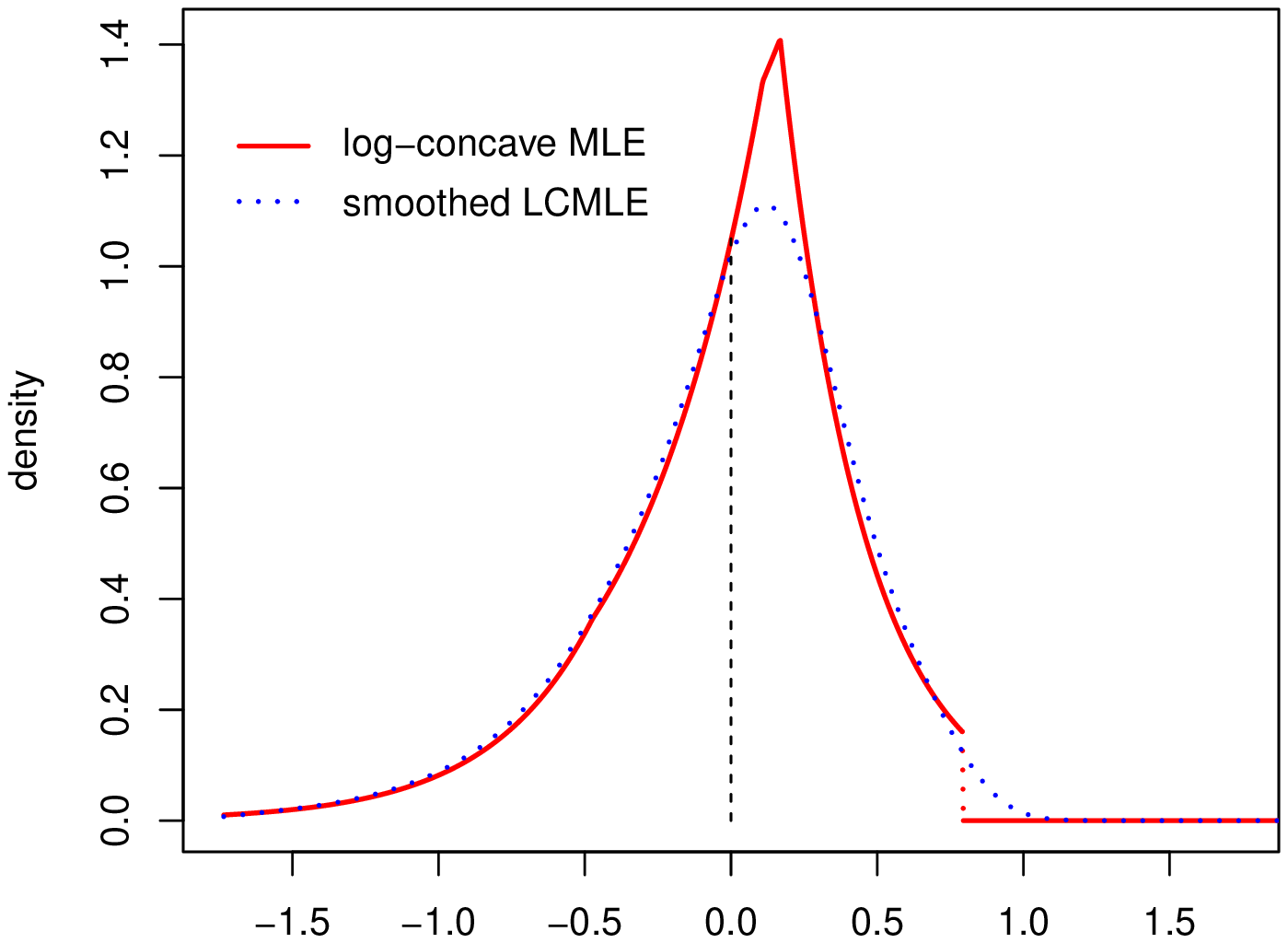} &
  \includegraphics[scale=0.4]{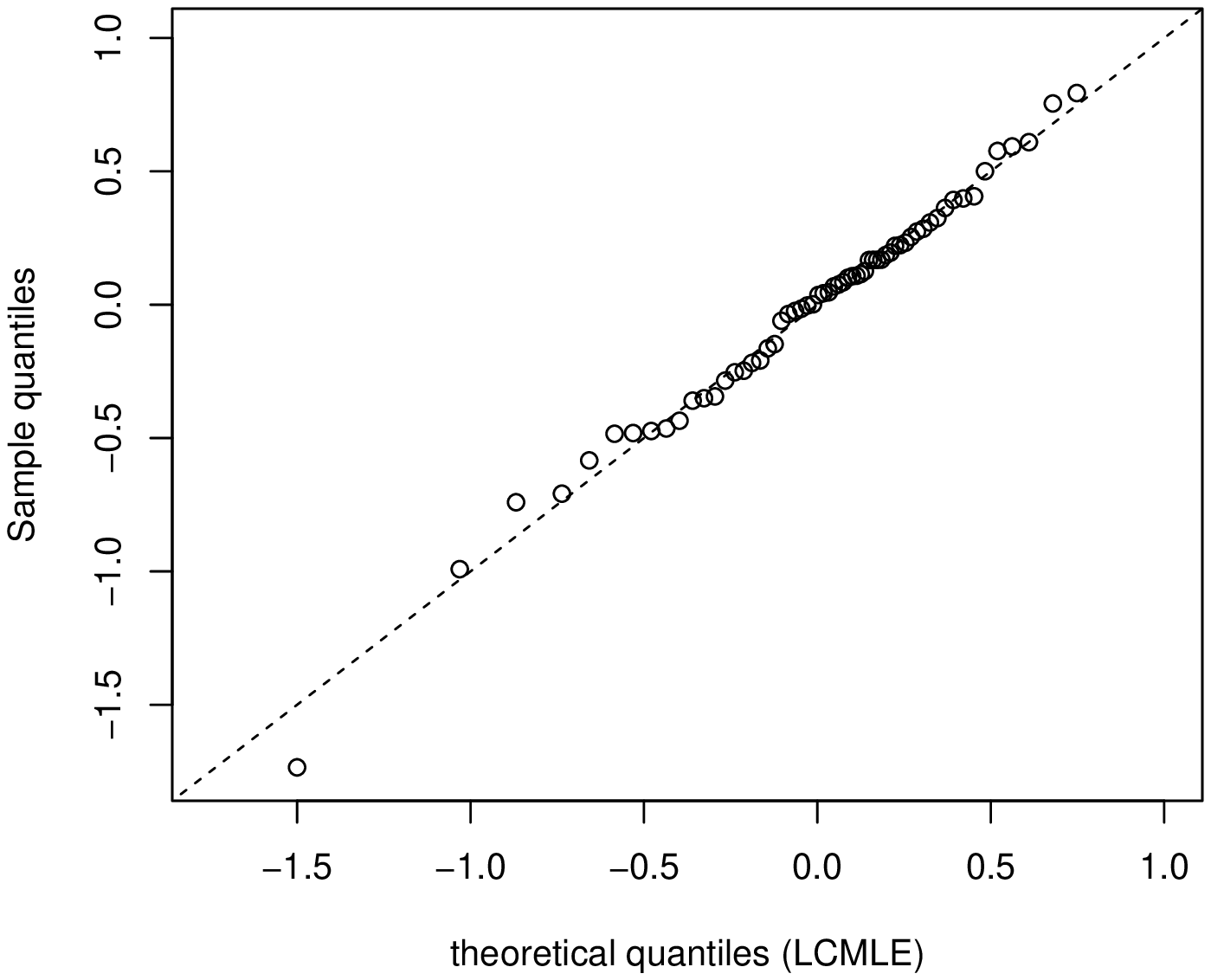} \\
  \mathrm{(c)} & \mathrm{(d)}
 \end{array}$
 \caption{(a) plots the log-transformed and centered time series based on the rabbit population data set; (b) plots the PACF; (c) plots the estimated density functions by the LCMLE (solid) and the smoothed LCMLE (dotted); (d) gives the Q-Q plot of the residuals against the distribution of the fitted unsmoothed LCMLE.}
 \label{Fig:rabbit}
\end{figure}

Here we illustrate the use of our method on the rabbit population data set of \citet{Middleton1934}, freely available at \emph{http://www.sw.ic.ac.uk/cpb/cpb/gpdd.html}. The numbers of rabbits killed yearly on a large estate in Yorkshire, England from 1867 to 1928 were recorded in this data set. Data were log-transformed and centered. This transformation is commonly used in population ecology thanks to the multiplicative nature of the population dynamics processes involving birth and death. Figure~\ref{Fig:rabbit}(a) shows the transformed series. Its partial autocorrelation function (PACF) is plotted in Figure~\ref{Fig:rabbit}(b). Note that the PACF is still a useful tool to help identify the appropriate order of AR($p$) processes even if $Q$ is non-Gaussian (see Theorem~8.1.2 of \citet{BrockwellDavis1991}). The PACF plot hints that we could summarize the series by a first-order autoregressive (AR(1)) model 
\[
	X_t = a X_{t-1} + \epsilon_t,
\]
where $\{\epsilon_t\}$ are \emph{i.i.d.} innovations following an unknown distribution $Q$. 

It can be shown that it is inadequate to summarize this series using AR(1) with Gaussian innovations. Actually, a Shapiro--Wilk test on the residuals gives strong evidence against the normality assumption (p-value = 0.0015). One alternative is to refit the model with innovations of other parametric forms, but one still has to choose the parametric family of the innovations beforehand. Here our approach offers a new possibility. By adapting the autoregressive models into our framework, we have fitted the AR(1) with $\hat{a}_{\mathrm{LCMLE}} = 0.5635$. The estimated density functions corresponding to both unsmoothed and smoothed LCMLE are plotted in Figure~\ref{Fig:rabbit}(c). A quantile-quantile (Q-Q) plot of the residuals (obtained from LCMLE) against the distribution of the fitted unsmoothed LCMLE is illustrated in Figure~\ref{Fig:rabbit}(d), indicating that the log-concavity assumption of $Q$ seems to be adequate here. The corresponding Q-Q plot against the fitted smoothed LCMLE appears to be similar, so is omitted for brevity.

\section*{Acknowledgments}
I am extremely grateful to my Ph.D supervisor, Richard Samworth, for suggesting this investigation and for many subsequent insightful conversations. I also owe thanks to Peter Craigmile and Bodhisattva Sen for their helpful suggestions. Finally, I would like to thank the associate editor and three anonymous reviewers for their valuable comments that help improve this manuscript substantially.

\section{Appendix}
%Our aim of this section is to prove Theorem~\ref{Thm:armaconsistency} and Theorem~\ref{Thm:armagarchconsistency}. We also mention a few intermediate results on log-concave approximation in this section, which may be of some independent interest.  For proofs of Theorem~\ref{Thm:armaexist}, Corollary~\ref{Cor:arexist}, Corollary~\ref{Cor:arconsistency} and Theorem~\ref{Thm:armagarchexist}, we refer to \citet{Chen2013}.

\subsection{Preliminaries}
We first introduce the $p^{\mathrm{th}}$ Mallows distance and the L\'{e}vy--Prokhorov distance as useful measures of distances between two probability distributions. The $p^{\mathrm{th}}$ Mallows distance is also known as the $p^{\mathrm{th}}$ Wasserstein distance. For historical reasons, when $p=1$, it is also called the Kantorovich--Rubinstein distance or the Earth Mover's distance. The L\'{e}vy--Prokhorov distance is a generalization of the L\'{e}vy metric defined in one dimension. 

More formally, for two probability measures $\mu$ and $\nu$ on the same Polish metric space equipped with the metric $d$, the $p^{\mathrm{th}}$ Mallows distance is defined as 
\[
    D_{p} (\mu, \nu) = \big[ \inf \mathbb{E} d( X , Y )^{p} \big]^{1/p},
\]
where the infimum is taken over all joint distributions of the random variables $X$ and $Y$ with marginals $\mu$ and $\nu$ respectively. 

The L\'{e}vy--Prokhorov distance is defined as 
\[
    D_{L} (\mu, \nu) = \inf \left\{ \epsilon > 0 | \mu(A) \leq \nu (A^{\epsilon}) + \epsilon \ \text{and} \ \nu (A) \leq \mu (A^{\epsilon}) + \epsilon ,\ \forall \text{ Borel sets } A \right\},
\]
where $A^{\epsilon}$ is the $\epsilon$-neighborhood of $A$.

Note that the L\'{e}vy--Prokhorov metric characterizes the topology of weak convergence. Furthermore, convergence with respect to any Mallows distance is slightly stronger than the weak convergence. See \citet{Villani2009} for a nice introduction to these topics.

Our next definition is useful in proving the theoretical properties of the LCMLE. 
Let $\mathcal{Q}$ be the family of all probability distributions on $\mathbb{R}$. Denote by $\mathcal{Q}^*$ the subset of $\mathcal{Q}$ which contains all distributions of finite expectation and non-zero variance. For $Q \in \mathcal{Q}$, define a profile log-likelihood type functional
\[
	L(Q) =  \sup_{\phi \in \Phi} \,  \left\{\int \phi \, dQ - \int e^{\phi(x)} dx + 1 \right\}.
\]
If $Q$ does not have finite expectation, $L(Q) = -\infty$. If $Q$ has zero variance, $L(Q) = \infty$. 

The above function $L(\cdot)$ is just a special (one-dimensional) case of what has been studied in \citet{DSS2011}. For the reader's convenience, we briefly recall some of their results which will turn to be useful in Section~\ref{Sec:proof}. The following three lemmas are respectively Theorem~2.2, Remarks~2.3-2.5 and Theorem~2.14-2.15 of \citet{DSS2011}.
\begin{lem}[Existence]
\label{Lem:DSSExist}
For all $Q \in \mathcal{Q}^*$, there exists a unique function 
\begin{align}
\label{Eq:lcdapprox}
	\psi(\cdot|Q) \in \argmax_{\phi \in \Phi}\, \left\{ \int \phi \, dQ - \int e^{\phi(x)} dx + 1 \right\}. 
\end{align}
Moreover, this function $\psi$ satisfies $\int e^{\psi(x)} dx = 1$ and 
\[
	\mathrm{int}(\mathrm{csupp}(Q)) \subseteq \mathrm{dom}(\psi) \subseteq \mathrm{csupp}(Q),
\]
where $\mathrm{int}, \mathrm{dom}, \mathrm{csupp}$ are interior, domain and convex support operators respectively. Here the convex support is defined as the smallest closed interval $[b_1,b_2]$ such that $Q([b_1,b_2]) = 1$. One may refer to \citet{Rockafellar1997} for the details of these definitions.
\end{lem}

\begin{lem}[Properties] Let $Q \in \mathcal{Q}^*$, then
\label{Lem:DSSProp}
\begin{enumerate}[(i)]
\setlength{\itemsep}{0pt}
\setlength{\parskip}{0pt}
\setlength{\parsep}{0pt}
\item \textbf{First moment equality}: $\int x e^{\psi(x|Q)} dx = \int x Q(dx)$.
\item \textbf{Affine equivariance}: for $a,b \in \mathbb{R}$ with $b \neq 0$, let $Q_{a,b}$ to be the distribution of $a+bX$ when $X$ has distribution $Q$, then $L(Q_{a,b}) = L(Q) - \log |b|$.
\item \textbf{Convexity}: $L(\cdot)$ is convex on $\mathcal{Q}^*$. More precisely, for any $Q_1,Q_2 \in \mathcal{Q}^*$ and $0 < t < 1$, 
$L(tQ_1+(1-t)Q_2) \le t L(Q_1) + (1-t)L(Q_2)$. The two sides are equal if and only if $\psi(\cdot|Q_1)=\psi(\cdot|Q_2)$.
\end{enumerate}
\end{lem}

\begin{lem}[Continuity] Let $Q \in \mathcal{Q}^*$ and $(Q_n)_n$ be a sequence of distributions in $\mathcal{Q}^*$.
\label{Lem:DSSCont}
\begin{enumerate}[(i)]
\setlength{\itemsep}{0pt}
\setlength{\parskip}{0pt}
\setlength{\parsep}{0pt}
\item If $\lim_{n \rightarrow \infty} D_L(Q_n,Q) = 0$, then $\limsup_{n \rightarrow \infty} L(Q_n) \le L(Q)$.
\item If $\lim_{n \rightarrow \infty} D_1(Q_n,Q) = 0$, then $\lim_{n \rightarrow \infty} L(Q_n) = L(Q)$. Moreover, the probability densities $f = e^{\psi(\cdot|Q)}$ and $f_n = e^{\psi(\cdot|Q_n)}$ satisfy $\lim_{n \rightarrow \infty} \int |f_n(x) - f(x)| dx = 0.$
\end{enumerate}
\end{lem}

\subsection{Proofs}
\label{Sec:proof}
\begin{prooftitle}{of Theorem~\ref{Thm:armaexist}}
First, we show that for any $n > p+q+1$, the following event is null:
\[
	\Omega = \{\exists \boldsymbol{\theta} \in \Theta, m \in \mathbb{R} \mbox{ s.t. } \tilde{\epsilon}_t (\boldsymbol{\theta}) = m, \ \mbox{ for }t = 1, \ldots, n\}.
\]
To do this, we need some well-known results from differential geometry. See \citet{GuilleminPollack1974} for background information.

For any set of fixed initial values, consider a function $H: \mathbb{R}^{2(p+q+1)} \rightarrow \mathbb{R}^{p+q+1}$ defined as follows:
\[
	H(\boldsymbol{\theta},m,X_1,\ldots,X_{p+q+1}) = (\tilde{\epsilon}_1(\boldsymbol{\theta})-m, \ldots,\tilde{\epsilon}_{p+q+1}(\boldsymbol{\theta})-m)^T.
\]
It is easy to check that $H$ is a smooth (i.e. $C^{\infty}$) function. Furthermore, the Jacobian matrix of $H$ has full-rank, because
\begin{align*}
	\mathrm{Rank}\left[ \frac{\partial H}{\partial \boldsymbol{\theta}} \left| \begin{array}{cccc} 
	\frac{\partial H_1}{\partial m} & \frac{\partial H_1}{\partial X_1} & \ldots & \frac{\partial H_1}{\partial X_{p+q+1}} \\
	\vdots & \vdots & & \vdots     \\
	\frac{\partial H_{p+q+1}}{\partial m} & \frac{\partial H_{p+q+1}}{\partial X_1} & \ldots & \frac{\partial H_{p+q+1}}{\partial X_{p+q+1}}
	\end{array} \right]\right.
	&=
	\mathrm{Rank}\left[ \frac{\partial H}{\partial \boldsymbol{\theta}} \left| \begin{array}{cccccc} 
	1 & 1 & & & & 0 \\
	1 & \frac{\partial H_2}{\partial X_1} & 1 & & &\\
	1 & \frac{\partial H_3}{\partial X_1} & \frac{\partial H_3}{\partial X_2} & \ddots & & \\
	\vdots & \vdots & \vdots & \ddots & \ddots & \\
	1 & \frac{\partial H_{p+q+1}}{\partial X_1} & \frac{\partial H_{p+q+1}}{\partial X_2} & \ldots & \frac{\partial H_{p+q+1}}{\partial X_{p+q}} & 1
	\end{array} \right]\right. \\
	& = p + q + 1.
\end{align*}
Therefore, $(0,\ldots,0)^T \in \mathbb{R}^{p+q+1}$ is a regular value of $H$.

Denote by $C \in \mathbb{R}^{p+q+1}$ the set in which for every $(X_1, \ldots, X_{p+q+1})^T \in C$, $(0,\ldots,0)^T \in \mathbb{R}^{p+q+1}$ is a critical value for $h_{X_1,\ldots,X_{p+q+1}}(\boldsymbol{\theta},m) = H(\boldsymbol{\theta},m,X_1,\ldots,X_{p+q+1})$. The transversality-density theorem \citep[page~216]{Fuente2000} shows that $C$ has Lebesgue measure zero. Since under assumption \textbf{(A.1)}, the distribution of $(X_1, \ldots, X_{p+q+1})^T$ has a probability density function, it is easy to check that $\mathbb{P}_{X_1, \ldots, X_{p+q+1}}(C) = 0$. Furthermore, for every vector $(X_1, \ldots, X_{p+q+1})^T$ on the complement of $C$, the vector $(0,\ldots,0)^T \in \mathbb{R}^{p+q+1}$ is regular for $h_{X_1,\ldots,X_{p+q+1}}(\boldsymbol{\theta},m)$. 

Now fix any $(X_1, \ldots, X_{p+q+1})^T \notin C$ and assume $\Omega$ holds. By the preimage theorem \citep[page~21]{GuilleminPollack1974}, the preimage $h_{X_1,\ldots,X_{p+q+1}}^{-1}((0,\ldots,0)^T)$ is a submanifold with zero dimension, thus contains at most countably many isolated points; consequently, conditioning on $\{X_t\}_{t=1}^{p+q+1}$, $X_{p+q+2}$ can only take values at countably many points. It follows from assumption \textbf{(A.1)} that the event $\Omega$ is null. 
 
Next, write
\[
	\Upsilon_n(\boldsymbol{\theta}) = \sup_{\phi \in \Phi} \Lambda_n(\phi,\boldsymbol{\theta}),
\]
where $\Lambda_n(\cdot,\cdot)$ is defined in (\ref{Eq:armamle1}). On the complement of $\Omega$, Lemma~\ref{Lem:DSSCont} entails the continuity of $\Upsilon_n(\cdot)$ over $\Theta$. This, combined with the compactness of $\Theta$, yields the existence of the LCMLE.
\hfill $\Box$
\end{prooftitle}
\vspace{0.5cm}

\begin{prooftitle}{of Corollary~\ref{Cor:arexist}}
In view of Theorem~\ref{Thm:armaexist}, it is enough to show that $\Upsilon_n(\boldsymbol{\theta})$ is coercive. One may refer to the proof of Corollary~\ref{Cor:arconsistency} for a similar argument.\hfill $\Box$
\end{prooftitle}
\vspace{0.5cm}

\begin{prooftitle}{of Theorem~\ref{Thm:armaconsistency}}
For any $\boldsymbol{\theta} \in \Theta$, denote by $\{\epsilon_t(\boldsymbol{\theta})\}$ the strictly stationary, ergodic and \emph{non-anticipative} solution of 
\begin{align}
\label{Eq:armasolution}
	\epsilon_t(\boldsymbol{\theta}) = X_t - \sum_{i=1}^{p} a_i X_{t-i} - \sum_{i=1}^{q} b_i \epsilon_{t-i}(\boldsymbol{\theta}), \ \forall t \in \mathbb{Z}.
\end{align}
Here by saying ``non-anticipative'', we mean a process which value at each time $t$ is a measurable function of the variables $X_{t-u}$, $u = 0,1,2,\ldots$.

Such solution exists because assumption \textbf{(A.4)} implies that all the ARMA processes with parameter vector in $\Theta$ are invertible, thus their innovations have AR($\infty$) representations, i.e., $\{\epsilon_t(\boldsymbol{\theta})\} = \frac{\mathbi{A}_{\boldsymbol{\theta}}(B)}{\mathbi{B}_{\boldsymbol{\theta}}(B)} X_t$, where $B$ is the backshift operator. In particular, $\{\epsilon_t(\boldsymbol{\theta}_0)\} = \{\epsilon_t\}$. See also \citet{BrockwellLindner2010} and \citet[page~204, Theorem~3]{Hannan1970}.

It is convenient to define the empirical innovation distributions as follows:
\[
	Q_{n,\boldsymbol{\theta}} = \frac{1}{n} \, \sum_{t=1}^n \delta_{\epsilon_t(\boldsymbol{\theta})} 
	\quad \mbox{ and } \quad 
	\tilde{Q}_{n,\boldsymbol{\theta}} = \frac{1}{n} \, \sum_{t=1}^n \delta_{\tilde{\epsilon}_t(\boldsymbol{\theta})} \, .
\]

Furthermore, let $\ldots,\mathring{X}_{-1},\mathring{X}_0, \mathring{X}_1,\ldots$ be an independent new realization of the existing ARMA($p,q$) process (i.e. with $Q_0$ and $\boldsymbol{\theta}_0$), and define $\{\mathring{\epsilon}_t(\boldsymbol{\theta})\}$ analogously as shown in (\ref{Eq:armasolution}). Denote the distribution of $\mathring{\epsilon}_1(\boldsymbol{\theta})$ by $Q_{\boldsymbol{\theta}}$. Note that $Q_{\boldsymbol{\theta}} = Q_0$.

We will establish our results in the following order:
\begin{enumerate}[(a)]
\setlength{\itemsep}{0pt}
\setlength{\parskip}{0pt}
\setlength{\parsep}{0pt}
\item $\lim_{n \rightarrow \infty} \sup_{\boldsymbol{\theta} \in \Theta} D_1 (Q_{n,\boldsymbol{\theta}}, \tilde{Q}_{n,\boldsymbol{\theta}}) = 0$, a.s., where $D_1$ is the $1^{\mathrm{st}}$ Mallows distance. 
\item $\liminf_{n \rightarrow \infty} \sup_{\Phi \times \Theta} \Lambda_n(\phi, \boldsymbol{\theta}) \ge L(Q_0)$, a.s.
\item $\lim_{n \rightarrow \infty} \sup_{\boldsymbol{\theta} \in \Theta} D_L (Q_{n,\boldsymbol{\theta}}, Q_{\boldsymbol{\theta}}) = 0$, a.s.
\item $\hat{\boldsymbol{\theta}}_n \rightarrow \boldsymbol{\theta}_0$, a.s. 
\item $\lim_{n \rightarrow \infty} \int \bigl| \hat{f}_n(x) - f_0^*(x) \bigr| \ dx = 0$, a.s.
\end{enumerate}

\textbf{(a) Asymptotic irrelevance of the initial values.} Rewrite (\ref{Eq:armasolution}) in matrix form 
\begin{align}
\label{Eq:armamatrix}
	\boldsymbol{\epsilon}_t(\boldsymbol{\theta}) = \mathbf{y}_t(\boldsymbol{\theta}) + M(\boldsymbol{\theta}) \boldsymbol{\epsilon}_{t-1}(\boldsymbol{\theta}), 
\end{align}
where
\[
	\boldsymbol{\epsilon}_t(\boldsymbol{\theta}) = 
	\left[ \begin{array}{c} 
	\epsilon_t(\boldsymbol{\theta}) \\ 
	\epsilon_{t-1}(\boldsymbol{\theta}) \\ 
	\vdots \\ 
	\epsilon_{t-q+1}(\boldsymbol{\theta})
	\end{array} \right], \quad
	\mathbi{y}_t(\boldsymbol{\theta}) = 
	\left[ \begin{array}{c}
	X_t - \sum_{i=1}^p a_i X_{t-i} \\ 0 \\ \vdots \\ 0
	\end{array} \right], \quad
	M(\boldsymbol{\theta}) = 
	\left[ \begin{array}{cccc} 
	-b_1 & -b_2 & \cdots & -b_q \\ 
	1 & 0 & \cdots & 0 \\ 
	\vdots & \ddots & & \vdots \\ 
	0 & \cdots & 1 & 0
	\end{array} \right].
\]
The spectral radius of a matrix $M$, denoted by $\rho(M)$, is defined as the greatest modulus of its eigenvalues. It is easy to check that under assumptions \textbf{(A.2)}, \textbf{(A.3)} and \textbf{(A.4)}
\begin{align}
\label{Eq:armaspectral}
	\sup_{\boldsymbol{\theta} \in \Theta} \rho(M(\boldsymbol{\theta})) < 1.
\end{align}
By iterating (\ref{Eq:armamatrix}), we have
\[
	\boldsymbol{\epsilon}_t(\boldsymbol{\theta}) = \mathbi{y}_t(\boldsymbol{\theta}) + M(\boldsymbol{\theta}) \mathbi{y}_{t-1}(\boldsymbol{\theta}) + \cdots + M^{t-1}(\boldsymbol{\theta}) \mathbi{y}_{1}(\boldsymbol{\theta}) + M^t(\boldsymbol{\theta}) \boldsymbol{\epsilon}_0(\boldsymbol{\theta}).
\]
Let $\tilde{\mathbi{y}}_t(\boldsymbol{\theta})$ be the vector obtained by replacing $X_0, \ldots,X_{1-p}$ with any fixed initial guesses. Let $\tilde{\boldsymbol{\epsilon}}_t(\boldsymbol{\theta})$ be the vector obtained by replacing $\epsilon_i(\boldsymbol{\theta})$ by $\tilde{\epsilon}_i(\boldsymbol{\theta})$ for all $i \le t$. We have
\[
	\tilde{\boldsymbol{\epsilon}}_t(\boldsymbol{\theta}) = \mathbi{y}_t(\boldsymbol{\theta}) + \sum_{i=1}^{t-p-1} M^i(\boldsymbol{\theta}) \mathbi{y}_{t-i}(\boldsymbol{\theta}) + M^{t-p}(\boldsymbol{\theta}) \tilde{\mathbi{y}}_{p}(\boldsymbol{\theta}) + \cdots + M^{t-1}(\boldsymbol{\theta}) \tilde{\mathbi{y}}_{1}(\boldsymbol{\theta}) + M^t(\boldsymbol{\theta}) \tilde{\boldsymbol{\epsilon}}_0(\boldsymbol{\theta}).	
\]
It follows immediately from (\ref{Eq:armaspectral}) that almost surely
\begin{align*}
	 \sup_{\boldsymbol{\theta} \in \Theta} |\tilde{\epsilon}_t(\boldsymbol{\theta}) - \epsilon_t(\boldsymbol{\theta})| &\le  \sup_{\boldsymbol{\theta} \in \Theta} \|\tilde{\boldsymbol{\epsilon}}_t(\boldsymbol{\theta}) - \boldsymbol{\epsilon}_t(\boldsymbol{\theta})\|_{2} \\
	& \le \sup_{\boldsymbol{\theta} \in \Theta} \left\| \sum_{i=1}^{\min(p,t)} M^{t-i}(\boldsymbol{\theta})(\tilde{\mathbi{y}}_i (\boldsymbol{\theta})- \mathbi{y}_i(\boldsymbol{\theta})) + M^t(\boldsymbol{\theta}) (\tilde{\boldsymbol{\epsilon}}_0(\boldsymbol{\theta}) - \boldsymbol{\epsilon}_0(\boldsymbol{\theta})) \right\|_{2} \le K \rho^t, \forall t \in \mathbb{N},
\end{align*}
where $K > 0$ and $0 < \rho < 1$ are two constants, and $\|\cdot\|_2$ is the Euclidean norm. Now elementary considerations show that almost surely 
\[
\limsup_{n \rightarrow \infty} \sup_{\boldsymbol{\theta} \in \Theta} D_1 (Q_{n,\boldsymbol{\theta}}, \tilde{Q}_{n,\boldsymbol{\theta}}) \le \limsup_{n \rightarrow \infty} \frac{1}{n} \sum_{t=1}^n K \rho^t = \limsup_{n \rightarrow \infty}\frac{1}{n}\frac{K}{1-\rho} = 0.
\]

\textbf{(b) The lower bound.} It is well known in the empirical process theory that $D_1(Q_{n,\boldsymbol{\theta}_0},Q_0) \stackrel{a.s.}{\rightarrow} 0$. This and point (a) entail $D_1(\tilde{Q}_{n,\boldsymbol{\theta}_0},Q_0) \stackrel{a.s.}{\rightarrow} 0$. By Lemma~\ref{Lem:DSSCont}, almost surely
\[
	\liminf_{n \rightarrow \infty} \sup_{\Phi \times \Theta} \Lambda_n(\phi, \boldsymbol{\theta}) \ge \liminf_{n \rightarrow \infty} \sup_{\phi \in \Phi} \Lambda_n(\phi, \boldsymbol{\theta}_0) = \liminf_{n \rightarrow \infty} L(\tilde{Q}_{n,\boldsymbol{\theta}_0})= L(Q_0),
\]
where $\Lambda_n(\cdot,\cdot)$ is given in (\ref{Eq:armamle1}).

\textbf{(c) Uniform convergence in $D_L$.} We combine a Prohorov type approach with the standard compactness argument to establish this point. For all $\boldsymbol{\theta} \in \Theta$ and any positive integer $k$, denote by $V_k(\boldsymbol{\theta})$ the open ball centered at $\boldsymbol{\theta}$ of radius $1/k$. 

We first show that for any fixed $\boldsymbol{\theta}^* \in \Theta$, almost surely
\begin{align}
\label{Eq:armalocalergodic1}
	\lim_{k \rightarrow \infty} \lim_{n \rightarrow \infty} \sup_{\boldsymbol{\theta} \in V_k(\boldsymbol{\theta}^*) \cap \Theta} D_{L} (Q_{n,\boldsymbol{\theta}}, Q_{\boldsymbol{\theta}^*}) = 0.
\end{align}
To see this, we note that for any fixed $u \in \mathbb{R}$, 
\begin{align*}
	\sup_{\boldsymbol{\theta} \in V_k(\boldsymbol{\theta}^*) \cap \Theta} \frac{1}{n} \sum_{t=1}^n \mathbf{1} \left\{ \epsilon_t(\boldsymbol{\theta}) \le u \right\} \le \frac{1}{n} \sum_{t=1}^n \sup_{\boldsymbol{\theta} \in V_k(\boldsymbol{\theta}^*) \cap \Theta} \mathbf{1} \left\{ \epsilon_t(\boldsymbol{\theta}) \le u \right\} \le \frac{1}{n} \sum_{t=1}^n \mathbf{1} \left\{ \inf_{\boldsymbol{\theta} \in V_k(\boldsymbol{\theta}^*) \cap \Theta}\epsilon_t(\boldsymbol{\theta}) \le u \right\}.
\end{align*}
Notice that the function $\mathbf{1} \left\{ \inf_{\boldsymbol{\theta} \in V_k(\boldsymbol{\theta}^*) \cap \Theta}\epsilon_t(\boldsymbol{\theta}) \le u \right\}$ is measurable because $\epsilon_t(\boldsymbol{\theta})$ is a continuous function. Therefore we can use Theorem~36.4 of \citet{Billingsley1995} and the pointwise ergodic theorem to deduce that almost surely
\[
	\limsup_{n \rightarrow \infty} \sup_{\boldsymbol{\theta} \in V_k(\boldsymbol{\theta}^*) \cap \Theta} \frac{1}{n} \sum_{t=1}^n \mathbf{1} \left\{ \epsilon_t(\boldsymbol{\theta}) \le u \right\} \le \mathbb{P} \left\{ \inf_{\boldsymbol{\theta} \in V_k(\boldsymbol{\theta}^*) \cap \Theta} \mathring{\epsilon}_1(\boldsymbol{\theta}) \le u \right\}.
\]
The monotone convergence theorem says that $ \mathbb{P} \left\{ \inf_{\boldsymbol{\theta} \in V_k(\boldsymbol{\theta}^*) \cap \Theta} \mathring{\epsilon}_1(\boldsymbol{\theta}) \le u \right\}$ decreases to $\mathbb{P} ( \mathring{\epsilon}_1(\boldsymbol{\theta}^*) \le u )$ as $k \rightarrow \infty$. Applying a similar argument to the infimum to obtain that almost surely
\begin{align}
\label{Eq:armalocalergodic2}
	\mathbb{P} (\mathring{\epsilon}_1(\boldsymbol{\theta}^*) < u) &\le \liminf_{k \rightarrow \infty} \liminf_{n \rightarrow \infty} \inf_{\boldsymbol{\theta} \in V_k(\boldsymbol{\theta}^*) \cap \Theta} \frac{1}{n} \sum_{t=1}^n \mathbf{1} \left\{ \epsilon_t(\boldsymbol{\theta}) \le u \right\}  \\
\label{Eq:armalocalergodic3}
	&\le \limsup_{k \rightarrow \infty} \limsup_{n \rightarrow \infty} \sup_{\boldsymbol{\theta} \in V_k(\boldsymbol{\theta}^*) \cap \Theta} \frac{1}{n} \sum_{t=1}^n \mathbf{1} \left\{ \epsilon_t(\boldsymbol{\theta}) \le u \right\} \le \mathbb{P} ( \mathring{\epsilon}_1(\boldsymbol{\theta}^*) \le u ).
\end{align}
The tightness of $\cup_{\boldsymbol{\theta} \in V_k(\boldsymbol{\theta}^*)} Q_{n,\boldsymbol{\theta}}$ then follows from (\ref{Eq:armalocalergodic2}) and (\ref{Eq:armalocalergodic3}) for sufficiently large $k$.

Now suppose (\ref{Eq:armalocalergodic1}) does not hold. Then it is possible to find a subsequence $k_j \in \mathbb{N}$ with $n(k_j) < n(k_{j+1})$ and $\boldsymbol{\theta}_{k_j} \in V_{k_{j}}(\boldsymbol{\theta}^*)$ for all $j \in \mathbb{N}$ such that
\[
	\lim_{j \rightarrow \infty} D_{L} (Q_{n(k_j),\boldsymbol{\theta}_{k_j}}, Q_{\boldsymbol{\theta}^*}) > 0.
\]
By the Prohorov's theorem, extracting a further subsequence if necessary, there exists a probability distribution $Q_*$ such that
\[
	\lim_{j \rightarrow \infty} D_{L} (Q_{n(k_j),\boldsymbol{\theta}_{k_j}}, Q_*) = 0.
\]
Therefore $D_{L}(Q_*, Q_{\boldsymbol{\theta}^*}) > 0$. An application of the Portmanteau theorem shows that there at least exists an $u \in \mathbb{R}$, such that
\[
	Q_{n(k_j),\boldsymbol{\theta}_{k_j}}((-\infty,u]) > Q_{\boldsymbol{\theta}^*}((-\infty,u]).
\]
But this contradicts (\ref{Eq:armalocalergodic3}) (using the fact that for any fixed $n$, $\sup_{\boldsymbol{\theta} \in V_k(\boldsymbol{\theta}^*) \cap \Theta} \frac{1}{n} \sum_{t=1}^n \mathbf{1} \left\{ \epsilon_t(\boldsymbol{\theta}) \le u \right\}$ is a decreasing function with respect to $k$). Consequently, (\ref{Eq:armalocalergodic1}) holds true.

Moreover, by a similar Prohorov type of argument, one can show that 
\begin{align}
\label{Eq:armaexpdiff}
	\lim_{k \rightarrow \infty} \lim_{n \rightarrow \infty} \sup_{\boldsymbol{\theta} \in V_k(\boldsymbol{\theta}^*) \cap \Theta} D_{L} (Q_{\boldsymbol{\theta}}, Q_{\boldsymbol{\theta}^*}) = 0.
\end{align}
Thus 
\[
	\lim_{k \rightarrow \infty} \lim_{n \rightarrow \infty} \sup_{\boldsymbol{\theta} \in V_k(\boldsymbol{\theta}^*) \cap \Theta} D_{L} (Q_{n, \boldsymbol{\theta}}, Q_{\boldsymbol{\theta}}) = 0, \ \mbox{a.s.}
\]
We conclude the proof of point (c) by a compactness argument. For any arbitrary $\delta > 0$, for every $\boldsymbol{\theta}^* \in \Theta$, we can find a neighborhood $V(\boldsymbol{\theta}^*)$ satisfying 
\[
	\limsup_{n \rightarrow \infty} \sup_{\boldsymbol{\theta} \in V(\boldsymbol{\theta}^*) \cap \Theta} D_{L} (Q_{n, \boldsymbol{\theta}}, Q_{\boldsymbol{\theta}})   \le \delta, \ \mbox{ a.s.}
\]
Because $\Theta$ is compact, there exists a finite subcover of $\Theta$ of the form $V(\boldsymbol{\theta}_1),\ldots,V(\boldsymbol{\theta}_k)$. Thus 
\begin{align*}
	\limsup_{n \rightarrow \infty} \sup_{\boldsymbol{\theta} \in \Theta} D_{L} (Q_{n, \boldsymbol{\theta}}, Q_{\boldsymbol{\theta}}) \le \limsup_{n \rightarrow \infty} \max_{j = 1,\ldots,k} \sup_{\boldsymbol{\theta} \in V(\boldsymbol{\theta}_j) \cap \Theta} D_{L} (Q_{n, \boldsymbol{\theta}}, Q_{\boldsymbol{\theta}}) \le \delta, \ \mbox{ a.s.}
\end{align*}
This completes the proof of point (c).

\textbf{(d) Convergence of $\hat{\boldsymbol{\theta}}_n$.} To verify the assertion it suffices to consider a sequence of \emph{fixed} observations $X_1,X_2, \ldots$ such that points (a) -- (c) hold true. Our proof relies on the  following simple result from analysis: assume that $\{m_n\}$ is a bounded sequence with the property that every convergent subsequence of $\{m_n\}$ converges to the same limit $m$, then $\{m_n\}$ must converge to $m$. Now consider any convergent subsequence of $\hat{\boldsymbol{\theta}}_n$ that converges to any arbitrary $\boldsymbol{\theta}^*$, which we denote by $\hat{\boldsymbol{\theta}}_{n(j)} \rightarrow \boldsymbol{\theta}^*$. Because $\Theta$ is compact, $\boldsymbol{\theta}^* \in \Theta$. Our goal is to show that $\boldsymbol{\theta}^* = \boldsymbol{\theta}_0$. Point (c), together with (\ref{Eq:armaexpdiff}), entails that
\[
	\lim_{j \rightarrow \infty} D_L(Q_{n(j),\hat{\boldsymbol{\theta}}_{n(j)}},Q_{\boldsymbol{\theta}^*}) = 0.
\]
Since the convergence in the Mallows metric $D_1$ is stronger than the weak convergence, combining this with point (a) leads to $\tilde{Q}_{n(j),\hat{\boldsymbol{\theta}}_{n(j)}} \stackrel{d}{\rightarrow} Q_{\boldsymbol{\theta}^*}$. Moreover, because $\mathring{\epsilon}_1(\boldsymbol{\theta}_0)$ and $\mathring{\epsilon}_1(\boldsymbol{\theta}^*)-\mathring{\epsilon}_1(\boldsymbol{\theta}_0)$ are independent, by Lemma~\ref{Lem:DSSCont} and Theorem~3.5 of \citet{DSS2011},
\[
	\limsup_{j \rightarrow \infty} L(\tilde{Q}_{n(j),\hat{\boldsymbol{\theta}}_{n(j)}}) \le L(Q_{\boldsymbol{\theta}^*}) \le L(Q_0).
\]
In light of point (b), this implies that there must exist a constant $m \in \mathbb{R}$ such that with probability one
\begin{align}
\label{Eq:armaequal}
	\mathring{\epsilon}_1(\boldsymbol{\theta}^*) - \mathring{\epsilon}_1(\boldsymbol{\theta}_0) = m. 
\end{align}
Let $B$ be the backshift operator. Under assumption \textbf{(A.4)}, $\mathbi{B}_{\boldsymbol{\theta}}(B)$ is invertible for all $\boldsymbol{\theta} \in \Theta$, so (\ref{Eq:armaequal}) is equivalent to 
\[
	\left\{ \frac{\mathbi{A}_{\boldsymbol{\theta}^*}(B)}{\mathbi{B}_{\boldsymbol{\theta}^*}(B)} - \frac{\mathbi{A}_{\boldsymbol{\theta}_0}(B)}{\mathbi{B}_{\boldsymbol{\theta}_0}(B)} \right\} \mathring{X}_1 = m, \ w.p.1.
\]
If the operator in $B$ on the left hand side was not null, then there would exist a constant linear combination of $\mathring{X}_{1},\mathring{X}_{0}, \mathring{X}_{-1},\ldots$. This is impossible since the innovations are nondegenerate by assumption \textbf{(A.1)} (or \textbf{(A.1*)}). Thus we have
\[
	\frac{\mathbi{A}_{\boldsymbol{\theta}^*}(z)}{\mathbi{B}_{\boldsymbol{\theta}^*}(z)} = \frac{\mathbi{A}_{\boldsymbol{\theta}_0}(z)}{\mathbi{B}_{\boldsymbol{\theta}_0}(z)}, \ \forall  |z| \le 1.
\]
It follows under assumption \textbf{(A.5)} that $\mathbi{A}_{\boldsymbol{\theta}^*} = \mathbi{A}_{\boldsymbol{\theta}_0}$ and $\mathbi{B}_{\boldsymbol{\theta}^*} = \mathbi{B}_{\boldsymbol{\theta}_0}$, so $\boldsymbol{\theta}^* = \boldsymbol{\theta}_0$.
Finally, since $\Theta$ is compact and the convergent subsequence is picked arbitrarily, we obtain $\hat{\boldsymbol{\theta}}_n \rightarrow \boldsymbol{\theta}_0$.

\textbf{(e) Convergence of $\hat{f}_n$.} Recall that the weak convergence of $Q_{n,\hat{\boldsymbol{\theta}}_n}$ to $Q_0$ is established in the proof of point (d). Denote by $\mu_k'(Q)$ the $k$-th moment of the distribution $Q$. We now show the convergence in the first moment, i.e. $\mu_1'(Q_{n,\hat{\boldsymbol{\theta}}_n}) \stackrel{a.s.}{\rightarrow} \mu_1'(Q_0)$. Using the notations from the proof of point (c) and applying the ergodic theorem to both the infimum and the supremum, we have that almost surely
\begin{align*}
	\liminf_{n \rightarrow \infty} \inf_{\boldsymbol{\theta} \in V_k(\boldsymbol{\theta}_0) \cap \Theta} \frac{1}{n} \sum_{t=1}^n \epsilon_t(\boldsymbol{\theta})  \ge  \mathbb{E} \inf_{\boldsymbol{\theta} \in V_k(\boldsymbol{\theta}_0) \cap \Theta} \mathring{\epsilon}_1(\boldsymbol{\theta}), \\
	\limsup_{n \rightarrow \infty} \sup_{\boldsymbol{\theta} \in V_k(\boldsymbol{\theta}_0) \cap \Theta} \frac{1}{n} \sum_{t=1}^n \epsilon_t(\boldsymbol{\theta})  \le  \mathbb{E} \sup_{\boldsymbol{\theta} \in V_k(\boldsymbol{\theta}_0) \cap \Theta} \mathring{\epsilon}_1(\boldsymbol{\theta}). 
\end{align*}
The continuity of $\mathring{\epsilon}_1(\boldsymbol{\theta})$ (with respect to $\boldsymbol{\theta}$) and the monotone convergence theorem entail that
\[
	\lim_{k \rightarrow \infty} \lim_{n \rightarrow \infty} \inf_{\boldsymbol{\theta} \in V_k(\boldsymbol{\theta}_0) \cap \Theta} \frac{1}{n} \sum_{t=1}^n \epsilon_t(\boldsymbol{\theta}) = \lim_{k \rightarrow \infty} \lim_{n \rightarrow \infty} \sup_{\boldsymbol{\theta} \in V_k(\boldsymbol{\theta}_0) \cap \Theta} \frac{1}{n} \sum_{t=1}^n \epsilon_t(\boldsymbol{\theta}) = \mathbb{E} \mathring{\epsilon}_1(\boldsymbol{\theta}_0), \ \mbox{a.s.}
\]
This, together with point (d), entails $\mu_1'(Q_{n,\hat{\boldsymbol{\theta}}_n}) \stackrel{a.s.}{\rightarrow} \mu_1'(Q_0)$. Now we can use Theorem~6.9 of \citet{Villani2009} to show almost sure convergence in the $1^{\mathrm{st}}$ Mallows metric of $Q_{n,\hat{\boldsymbol{\theta}}_n}$ to $Q_0$. Moreover, it follows from point (a) that $D_1(\tilde{Q}_{n,\hat{\boldsymbol{\theta}}_n}, Q_0) \stackrel{a.s.}{\rightarrow} 0$. Point (e) can now be established via Lemma~\ref{Lem:DSSCont}.
\hfill $\Box$
\end{prooftitle}
\vspace{0.5cm}

\begin{prooftitle}{of Corollary~\ref{Cor:arconsistency}}
In view of the proof of Theorem~\ref{Thm:armaconsistency}, all that remains is to show the almost sure boundedness of $\|\hat{\boldsymbol{\theta}}_n\|_2$. Let $\mu_X = \mathbb{E} \mathring{X}_0 = \frac{\int x f_0(x) dx}{\mathbi{A}_{\boldsymbol{\theta}_0}(1)}$. Using the fact that $\tilde{\epsilon}_t(\hat{\boldsymbol{\theta}}_n) = \epsilon_t + \sum_{i=1}^p (a_{0i}-\hat{a}_{ni}) X_{t-i}$ and with some careful calculations, we have 
\begin{align*}
	\int |t - \mu_1'(Q_{n,\hat{\boldsymbol{\theta}}_n})| Q_{n,\hat{\boldsymbol{\theta}}_n}(dt) &\ge \frac{1}{n} \sum_{t=1}^{n} \left| \sum_{i=1}^p (a_{0i}-\hat{a}_{ni}) (X_{t-i} - \mu_X) \right|\\
	&- \left| \frac{1}{n} \sum_{t=1}^{n} \sum_{i=1}^p (a_{0i}-\hat{a}_{ni}) (X_{t-i}- \mu_X) \right| 
	- \frac{1}{n}\sum_{t=1}^{n}|\epsilon_t| - \left|\frac{1}{n}\sum_{t=1}^{n}\epsilon_t \right|.
\end{align*}
It follows from Lemma~3.1 of \citet{DSS2011}, the law of large numbers and point (b) in the previous proof that
\begin{align}
\label{Eq:arupperbound}
	\frac{1}{n} \sum_{t=1}^{n} \left|\sum_{i=1}^p (a_{0i}-\hat{a}_{ni}) (X_{t-i}-\mu_X)\right| - \frac{1}{n} \left|\sum_{t=1}^{n} \sum_{i=1}^p (a_{0i}-\hat{a}_{ni}) (X_{t-i}-\mu_X)\right| < C_1
\end{align}
almost surely, for sufficiently large $n \in \mathbb{N}$, provided that $C_1 >  2 \int |t| f_0(dt) + e^{-L(Q_0)}$.

Let's consider the set $\{\boldsymbol{\theta} \in \mathbb{R}^p: \|\boldsymbol{\theta} - \boldsymbol{\theta}_0\|_2 = 1\}$. By the uniform ergodic theorem, almost surely 
\begin{align}
\label{Eq:aruniformergodic1}
	\lim_{n \rightarrow \infty} \sup_{\boldsymbol{\theta}: \|\boldsymbol{\theta} - \boldsymbol{\theta}_0\|_2 = 1} \left| \frac{1}{n} \sum_{t=1}^{n} \left|\sum_{i=1}^p (a_{0i}- a_i) (X_{t-i}-\mu_X)\right| - \mathbb{E}\left| \sum_{i=1}^p (a_{0i}- a_i) (\mathring{X}_{p+1-i}-\mu_X)\right| \right| &= 0,\\
\label{Eq:aruniformergodic2} 
	\lim_{n \rightarrow \infty} \sup_{\boldsymbol{\theta}: \|\boldsymbol{\theta} - \boldsymbol{\theta}_0\|_2 = 1} \left| \frac{1}{n} \sum_{t=1}^{n} \sum_{i=1}^p (a_{0i}- a_i) (X_{t-i}-\mu_X)\right| &= 0.
\end{align}

Observe that $\mathbb{E} |\sum_{i=1}^p (a_{0i}-a_i) (\mathring{X}_{p+1-i}-\mu_X)| > 0$, because otherwise $\{\mathring{X}_1-\mu_X, \ldots, \mathring{X}_p-\mu_X \}$ would be linearly dependent, which would violate assumption \textbf{(A.1)} or \textbf{(A.1*)}. By the compactness of $\{\boldsymbol{\theta} \in \mathbb{R}^p: \|\boldsymbol{\theta} - \boldsymbol{\theta}_0\|_2 = 1\}$, 
\[ 
	\min_{\boldsymbol{\theta}: \|\boldsymbol{\theta} - \boldsymbol{\theta}_0 \|_2 = 1} \mathbb{E}\left|\sum_{i=1}^p(a_{0i}-a_i)(\mathring{X}_{p+1-i}-\mu_X)\right| = C_2 > 0.
\]
Because of the scaling property, 
\begin{align}
\label{Eq:arlowerbound}
	\min_{\boldsymbol{\theta}: \|\boldsymbol{\theta} - \boldsymbol{\theta}_0 \|_2  = u} \mathbb{E}\left|\sum_{i=1}^p(a_{0i}-a_i)(\mathring{X}_{p+1-i}-\mu_X)\right| = u C_2.
\end{align}
Putting (\ref{Eq:arupperbound}), (\ref{Eq:aruniformergodic1}), (\ref{Eq:aruniformergodic2}) and (\ref{Eq:arlowerbound}) together entails that almost surely $\|\hat{\boldsymbol{\theta}}_n - \boldsymbol{\theta}_0\|_2 \le C_1/C_2 $, which also implies that $\|\hat{\boldsymbol{\theta}}_n\|_2$ is bounded.
\hfill $\Box$
\end{prooftitle}
\vspace{0.5cm}

\begin{prooftitle}{of Theorem~\ref{Thm:armagarchexist}}
Following the scheme of the proof of Theorem~\ref{Thm:armaexist}, it suffices to show that for $n > p + q + r + s + 1$ the following event is null:
\[
	\Omega = \left\{\exists \boldsymbol{\theta} \in \Theta', m \in \mathbb{R} \mbox{ s.t. } \tilde{\eta}_{t}(\boldsymbol{\theta}) = m \tilde{\sigma}_{t}(\boldsymbol{\theta}), \ \mbox{ for }t = 1, \ldots, n \right\}.
\]

Now let's construct the function $H: \Theta' \times \mathbb{R} \rightarrow \mathbb{R}^{p+q+r+s+1}$ as 
\[
	H(\boldsymbol{\theta},m,X_1,\ldots,X_{p+q+r+s+1}) = (\tilde{\eta}_{1}(\boldsymbol{\theta}) - m \tilde{\sigma}_{1}(\boldsymbol{\theta}), \ldots, \tilde{\eta}_{p+q+r+s+1}(\boldsymbol{\theta}) - m \tilde{\sigma}_{p+q+r+s+1}(\boldsymbol{\theta}))^T.
\]
Note that $H$ is actually a $\mathbb{R}^{2(p+q+r+s+1)} \rightarrow \mathbb{R}^{p+q+r+s+1}$ mapping, because the $(p+q+1)^{\mathrm{th}}$ component of $\Theta'$ is always  one.

The rest of the proof is similar to that of Theorem~\ref{Thm:armaexist}, so is omitted.
\hfill $\Box$
\end{prooftitle}
\vspace{0.5cm}

Before proceeding to prove Theorem~\ref{Thm:armagarchconsistency}, we establish a few useful intermediate results. The following lemma is a version of Slutsky's theorem with respect to the $1^{\mathrm{st}}$ Mallows distance.
\begin{lem}
\label{Lem:d1conv}
Let $X_0, X_1, X_2, \ldots$ be univariate random variables with corresponding distributions $P_0, P_1, P_2, \ldots$. Suppose $\mathbb{E}|X_0|< \infty$ and $D_1(P_n,P_0) \rightarrow 0$.
\begin{enumerate}[(i)]
\setlength{\itemsep}{0pt}
\setlength{\parskip}{0pt}
\setlength{\parsep}{0pt}
\item Let $m_1,m_2, \ldots$ be a real sequence with finite limit $\lim_{n \rightarrow \infty} m_n = m_0$. Denote by $Q_0, Q_1,\ldots$ the corresponding distributions of $m_0 X_0, m_1 X_1, \ldots$, then $D_1(Q_n,Q_0) \rightarrow 0$.
\item Let $Y$ be a univariate random variable independent of $\{X_i\}_{i=0}^{\infty}$ with $\mathbb{E}|Y| < \infty$. Denote by $Q_0,Q_1,\ldots$ the corresponding distributions of $X_0 Y, X_1 Y, \ldots$, then $D_1(Q_n,Q_0) \rightarrow 0$.
\end{enumerate}
\end{lem}

\begin{prooftitle}{of Lemma~\ref{Lem:d1conv}}
We only show (i) here. One can use a similar argument to prove (ii). 

Recall that the definition of the $1^{\mathrm{st}}$ Mallows distance is $D_1(Q_n, Q_0) = \inf_{(X_n,X_0)} \mathbb{E} |m_n X_n - m_0 X_0|$,
where the infimum is taken over all pairs $(X_n, X_0)$ of random variables $X_n \sim P_n$, $X_0 \sim P_0$ on a common probability space.
Since $D_1$ convergence implies $\mathbb{E} |X_n|  \rightarrow \mathbb{E}|X_0| < \infty$, we have
\begin{align*}
	\inf_{(X_n,X_0)} \mathbb{E} |m_n X_n - m_0 X_0| &\le \inf_{(X_n, X_0)} \big\{ \mathbb{E} |m_n X_n - m_0 X_n| + \mathbb{E} |m_0 X_n - m_0 X_0| \big\} \\
	& \le |m_n - m_0| \, \mathbb{E} |X_n| + m_0 \inf_{(X_n, X_0)} \mathbb{E} |X_n - X_0| \ {\rightarrow} \ 0 ,
\end{align*}
as desired.
\hfill $\Box$
\end{prooftitle}
\vspace{0.5cm}

The next lemma enhances our understanding of the behavior of the functional $\psi(\cdot|Q)$ given in (\ref{Eq:lcdapprox}). 
\begin{lem}
\label{Lem:lcdpsi}
Let $X_u, X_l, Y$ be univariate random variables. Let $R_u$, $R_l$ and $Q$ be the corresponding distributions of $X_uY$, $X_lY$ and $Y$. Assume that
\begin{enumerate}[(i)]
\setlength{\itemsep}{0pt}
\setlength{\parskip}{0pt}
\setlength{\parsep}{0pt}
\item $X_u$ and $Y$ are independent, with $\mathbb{E}|X_u|< \infty$;
\item $X_l$ and $Y$ are independent;
\item $Q \in \mathcal{Q}^*$;
\item There exists $m > 0$ such that $\mathbb{P}(X_u > m) = 1$ and $\mathbb{P}(m \ge X_l > 0) = 1$.
\end{enumerate}
Then $\psi(\cdot|R_u) \ne \psi(\cdot|R_l)$.
\end{lem}
\begin{prooftitle}{of Lemma~\ref{Lem:lcdpsi}}
First we show that both $\psi(\cdot|R_u)$ and $\psi(\cdot|R_l)$ uniquely exist. In view of Lemma~\ref{Lem:DSSExist}, it is enough to check that $R_u \in \mathcal{Q}^*$ and $R_l \in \mathcal{Q}^*$. This can be easily done using the facts that $Q \in \mathcal{Q}^*$, $\mathbb{E}|X_u|< \infty$ and $\mathbb{E}|X_l|< \infty$.

Now suppose $\psi(\cdot|R_u) = \psi(\cdot|R_l) = \psi(\cdot)$. We claim that the expectation of $Y$ is zero. This is due to the first moment equality in Lemma~\ref{Lem:DSSProp}. Moreover,  the convex support of $Q$ must be $\mathbb{R}$. Otherwise, by the second part of Lemma~\ref{Lem:DSSExist}, the domains of $\psi(\cdot|R_u)$ and $\psi(\cdot|R_l)$ would be different, which would contradict $\psi(\cdot|R_u) = \psi(\cdot|R_l)$.

Because $\psi(\cdot)$ is concave and $e^\psi$ defines a density, there exists $v \in (-\infty, \infty)$ such that 
\[
	\psi(v) > \frac{1}{2} \left\{ \psi(v-\delta)+\psi(v+\delta) \right\} \mbox{ for all } \delta > 0. 
\]
Without loss of generality, we may assume $v \le 0$, since otherwise by symmetry one may just take the additive inverse of $Y$.

Let $G$ be the cumulative distribution function with log-density $\psi$. Then by Theorem~2.7 of \citet{DSS2011},
\begin{align*}
	\int_{-\infty}^v \left\{\mathbb{P}(X_uY \le t) - G(t)\right\}\,dt = 0 \quad \mbox{ and } \quad \int_{-\infty}^v \left\{\mathbb{P}(X_lY \le t) - G(t)\right\}\,dt = 0  .
\end{align*}
It follows that 
\begin{align}
\label{Eq:lcdpsi1}
	\int_{-\infty}^v \left\{\mathbb{P}(X_uY \le t) - \mathbb{P}(X_lY \le t)\right\}\,dt = 0.
\end{align}
Note that for every $t \in (-\infty, v] \subseteq (-\infty, 0]$, we have 
\begin{align}
\label{Eq:lcdpsi2}
	\mathbb{P}(X_lY \le t) \le \mathbb{P}(Y \le t/m)\le \mathbb{P}(X_uY \le t). 
\end{align}
Because cumulative distribution functions are right continuous with left limits (c\'{a}dl\`{a}g), (\ref{Eq:lcdpsi1}) and (\ref{Eq:lcdpsi2}) imply that 
\[
	\mathbb{P}(X_uY \le t) = \mathbb{P}(Y \le t/m) = \mathbb{P}(X_lY \le t), \, \mbox{ for every } t \in (-\infty, v). 
\]

As $\mathbb{P}(X_u > m) = 1$, we can find some $\delta > 0$ such that $\mathbb{P}(X_u > m + \delta) > 0$. Now
\[
	\mathbb{P}(Y \le t/m) = \mathbb{P}(X_uY \le t) \ge \mathbb{P}(X_u > m + \delta) \mathbb{P}\left(Y \le \frac{t}{m+\delta}\right) + \mathbb{P}(m+\delta \ge X_u > m ) \mathbb{P}(Y \le t/m).
\]
From above, we obtain $\mathbb{P}(Y \le t/m) \ge \mathbb{P}\left(Y \le \frac{t}{m+\delta}\right)$, which implies $\mathbb{P}(Y \le t/m) = \mathbb{P}\left(Y \le \frac{t}{m+\delta}\right)$ for all $t \in (-\infty, v) \subseteq (-\infty,0)$.
Consequently, if we take any fixed $t \in (-\infty,v)$, then
\[
	\mathbb{P}(Y \le t/m) = \sum_{i=1}^{\infty} \mathbb{P}\left\{\frac{t}{m}\left(\frac{m+\delta}{m}\right)^i  < Y \le \frac{t}{m}\left(\frac{m+\delta}{m}\right)^{i-1}\right\} = 0.
\]

On the other hand, because the convex support of $Q$ is $\mathbb{R}$, we must have $\mathbb{P}(Y \le t/m) > 0$ for every $t < 0$. The proof is complete by Reductio ad absurdum.   
\hfill $\Box$
\end{prooftitle}
\vspace{0.5cm}

The following theorem can be viewed as a version of Jensen's inequality on $\mathcal{Q}^*$. It serves as the key ingredient in proving Theorem~\ref{Thm:armagarchconsistency}.
\begin{thm}
\label{Thm:timesnoisel}
Let $X, Y$ be univariate random variables with corresponding distributions $P, Q$ and $Q \in \mathcal{Q}^*$. Suppose further that $X$ and $Y$ are independent,  with $\mathbb{P}(X \ge 0) = 1$ and $\mathbb{E}\log X = m < \infty$. Denote the distribution of $XY$ by $R$. Then
\begin{align}
\label{Eq:timesnoisel}
	L(R) \le L(Q) - m.
\end{align}
The equality holds if and only if $X=e^m$ with probability one.
\end{thm}

\begin{prooftitle}{of Theorem~\ref{Thm:timesnoisel}}
The inequality is trivial in the following cases:
\begin{enumerate}[(i)]
\setlength{\itemsep}{0pt}
\setlength{\parskip}{0pt}
\setlength{\parsep}{0pt}
\item $\mathbb{E} X = \infty$\,: Because $Q \in \mathcal{Q}^*$, $\mathbb{E}|Y| > 0$ and $L(Q)$ is finite. Note that $\mathbb{E}|XY| = \mathbb{E} |X| \,\mathbb{E}|Y| = \infty$, so $L(R) = -\infty$.  In this case, the inequality ({Eq:timesnoisel}) is strict.
\item $\mathrm{var}(X) = 0$\,: $P$ is a point mass, so $L(R) = L(Q) - m$ by the affine equivariance of $L(\cdot)$.
\item $\mathbb{E}\log X = -\infty$\,: The right hand side of ({Eq:timesnoisel}) is $\infty$, so the inequality always holds. Now for the equality to hold, one needs $L(R) = \infty$, thus $R$ is a point mass. It then follows that $\mathbb{P}(X=0)=1$.
\end{enumerate}
For the remaining of the proof, we assume $P \in \mathcal{Q}^*$ and $m > -\infty$. It is implied that $R \in \mathcal{Q}^*$.

Denote by $F$ and $G$ the cumulative distribution functions corresponding to $P$ and $Q$. Let $X_n$ be a random variable independent of $Y$ and with the corresponding distribution $P_n$ defined as
\[
	P_n = \frac{1}{n} \sum_{i=1}^{n}\delta_{F^{-1}(\frac{i}{n+1})} \ ,
\]
where $F^{-1}$ is the generalized inverse function of $F$, i.e. $F^{-1}(p)= \inf\{ x\in \mathbb{R} : p \le F(x) \}$. In other words, $X_n$ is the ``stratified'' approximation of $X$.

Let $R_n$ be the distribution corresponding to $X_n Y$. Abusing notation slightly in the following, given $t \in \mathbb{R}$, we denote $Q_t$ to be the distribution corresponding to the random variable $tY$. Then $R_n = \frac{1}{n} \sum_{i=1}^{n} Q_{F^{-1}(\frac{i}{n+1})}$. Because $L(\cdot)$ is convex and affine equivariant (Lemma~\ref{Lem:DSSProp}),
\begin{align}
\label{Eq:lconvex}
	L(R_n) \le \frac{1}{n} \sum_{i=1}^{n} L(Q_{F^{-1}(\frac{i}{n+1})}) = L(Q) - \frac{1}{n} \sum_{i=1}^{n} \log F^{-1}\left(\frac{i}{n+1}\right).
\end{align}

Since $D_1(P_n,P) \rightarrow 0$, Lemma~\ref{Lem:d1conv}(ii) shows that $D_1(R_n,R) \rightarrow 0$. It follows from Lemma~\ref{Lem:DSSCont} that $\lim_{n \rightarrow \infty} L(R_n) = L(R)$. Furthermore, 
\[
	\lim_{n \rightarrow \infty} \frac{1}{n} \sum_{i=1}^{n} \log F^{-1}\left(\frac{i}{n+1}\right) = \int_0^1 \log F^{-1}(p) dp = m.
\]
We now let $n \rightarrow \infty$ on both sides of (\ref{Eq:lconvex}) to establish the inequality (\ref{Eq:timesnoisel}).

Next, we show that (\ref{Eq:timesnoisel}) is strict if $P \in \mathcal{Q}^*$. Fix $v = F^{-1}(1/2)$. It follows from $m > -\infty$ that $v > 0$ and $\mathbb{P}(X > 0) = 1$. Since we have assumed that $X$ is not almost surely constant (i.e. $\mathrm{var}(X) > 0$), $\mathbb{P}(X \ge v) = p \in [1/2,1)$. Denote by $R_u$ and $R_l$ the corresponding distributions of $(XY|X \ge v)$ and $(XY|X < v)$. Clearly, $R = p R_u + (1-p) R_l$. From Lemma~\ref{Lem:lcdpsi}, $\psi(\cdot|R_u) \ne \psi(\cdot|R_l)$. Now by the convexity of $L(\cdot)$ (Lemma~\ref{Lem:DSSProp}(iii)) again, we have
\[
	L(R) < p L(R_u) + (1-p) L(R_l).
\]
Using the inequality part of (\ref{Eq:timesnoisel}) proved above,
\begin{align*}
	p L(R_u) + (1-p) L(R_l) &\le p L(Q) - \mathbb{E}(\log X \mathbf{1}\{X \ge v\}) + (1-p) L(Q) -  \mathbb{E}(\log X \mathbf{1}\{X < v\}) \\
	& = L(Q) - \mathbb{E}\log X = L(Q) - m.
\end{align*}
Consequently, $L(R) < L(Q) - m$, as required. 
\hfill $\Box$
\end{prooftitle}
\vspace{0.5cm}

The next corollary is combination of Theorem~3.5 of \citet{DSS2011} and our Theorem~\ref{Thm:timesnoisel}. Its proof is omitted owing to its similarity to that of Theorem~\ref{Thm:timesnoisel}.
\begin{cor}
\label{Cor:addtimesnoisel}
Let $X_1,X_2, Y$ be univariate random variables with corresponding distributions $P_1, P_2$ and $Q$. $Q \in \mathcal{Q}^*$. Suppose that $X_1$ and $Y$ are independent, $X_2$ and $Y$ are independent,  with $\mathbb{P}(X_2 \ge 0) = 1$ and $\mathbb{E}\log X_2 = m \in (-\infty, \infty)$. Denote the distribution of $(X_1+Y)X_2$ by $R$. Then
\[
	L(R) \le L(Q) - m.
\]
The equality holds if and only if $P_1 = \delta_u$ for some $u \in \mathbb{R}$ and $P_2=\delta_{e^m}$.
\end{cor}

\begin{prooftitle}{of Theorem~\ref{Thm:armagarchconsistency}}
Under assumptions \textbf{(A.4)} and \textbf{(B.4)}, $\{X_t\}$ is stationary and ergodic. Let $\{\eta_t(\boldsymbol{\theta})\}$ and $\{\sigma_t^2(\boldsymbol{\theta})\}$ be respectively the stationary, ergodic and non-anticipative solutions of 
\begin{align}
\label{Eq:armagarchsolution1}
	\eta_t(\boldsymbol{\theta}) &= X_t - \sum_{i=1}^{p} a_i X_{t-i} - \sum_{i=1}^{q} b_i \eta_{t-i}(\boldsymbol{\theta}), \ \forall t \in \mathbb{Z}, \\
\label{Eq:armagarchsolution2}
	\sigma_t^2(\boldsymbol{\theta}) &= c + \sum_{i=1}^{r} \alpha_{i} \eta_{t-i}^2(\boldsymbol{\theta}) + \sum_{i=1}^{s} \beta_{i} \sigma_{t-i}^2(\boldsymbol{\theta}), \ \forall t \in \mathbb{Z}.
\end{align}
Note that assumptions \textbf{(A.4)} and \textbf{(B.2)}--\textbf{(B.4)} ensure the existence of such solutions.

Define the empirical distributions as 
\begin{align*}
	Q_{n,\boldsymbol{\theta}} & = \frac{1}{n} \, \sum_{t=1}^n \delta_{\eta_t(\boldsymbol{\theta})/ \sigma_t(\boldsymbol{\theta})} \quad  \mbox{ and } \quad
	\tilde{Q}_{n,\boldsymbol{\theta}} = \frac{1}{n} \, \sum_{t=1}^n \delta_{\tilde{\eta}_t(\boldsymbol{\theta})/\tilde{\sigma}_t(\boldsymbol{\theta})} \ .
\end{align*}

Let $\ldots,\mathring{X}_{-1}, \mathring{X}_0, \mathring{X}_1, \ldots$ be an independent new realization of the existing ARMA($p,q$)-GARCH($r,s$), and define $\{\mathring{\eta}_t(\boldsymbol{\theta})\}$ and $\{\mathring{\sigma}_t^2(\boldsymbol{\theta})\}$ analogously as shown in (\ref{Eq:armagarchsolution1}) and (\ref{Eq:armagarchsolution2}). Denote the distribution of $\frac{\mathring{\eta}_1(\boldsymbol{\theta})}{\mathring{\sigma}_1(\boldsymbol{\theta})}$ by $Q_{\boldsymbol{\theta}}$.

We will split our proof into several parts:
\begin{enumerate}[(a)]
\setlength{\itemsep}{0pt}
\setlength{\parskip}{0pt}
\setlength{\parsep}{0pt}
\item $\lim_{n \rightarrow \infty} \sup_{\boldsymbol{\theta} \in \Theta'} D_2 (Q_{n,\boldsymbol{\theta}}, \tilde{Q}_{n,\boldsymbol{\theta}}) = 0$, a.s., where $D_2$ is the $2^{\mathrm{nd}}$ Mallows distance. 
\item $\lim_{n \rightarrow \infty} \sup_{\boldsymbol{\theta} \in \Theta'} \frac{1}{2n} \left|\sum_{t=1}^n \log \tilde{\sigma}_t^2(\boldsymbol{\theta}) - \sum_{t=1}^n \log \sigma_t^2(\boldsymbol{\theta}) \right| = 0$, a.s.
\item For any $\boldsymbol{\theta} \in \Theta'$, $\mathbb{E} \log \mathring{\sigma}_1^2(\boldsymbol{\theta}) < \infty$.
\item $\liminf_{n \rightarrow \infty} \sup_{\Phi \times \Theta'} \Lambda_n(\phi, \boldsymbol{\theta}) \ge L(Q_0) - \frac{1}{2} \mathbb{E} \log \mathring{\sigma}_1^2(\boldsymbol{\theta}_0)$, a.s.
\item $\lim_{n \rightarrow \infty} \sup_{\boldsymbol{\theta} \in \Theta'} \left|\frac{1}{n} \sum_{t=1}^n \log \sigma_t^2(\boldsymbol{\theta}) - \mathbb{E} \log \mathring{\sigma}_1^2(\boldsymbol{\theta}) \right| = 0$, a.s.
\item $\lim_{n \rightarrow \infty} \sup_{\boldsymbol{\theta} \in \Theta'} D_L (Q_{n,\boldsymbol{\theta}}, Q_{\boldsymbol{\theta}}) = 0$, a.s.
\item $\hat{\boldsymbol{\theta}}'_n \rightarrow \boldsymbol{\theta}_0'$, a.s., where we write for convenience
\[
	\boldsymbol{\theta}_0' = \left(a_{01},\ldots,a_{0p},b_{01},\ldots,b_{0q}, 1,\frac{\alpha_{01}}{c_0},\ldots,\frac{\alpha_{0r}}{c_0},\beta_{01},\ldots,\beta_{0s}\right)^T .
\]
\item $\hat{c}_n \rightarrow c_0$, a.s.
\item $\lim_{n \rightarrow \infty} \int \bigl| \hat{f}_n(x) - f_0^*(x) \bigr| dx = 0$, a.s.
\end{enumerate}

%\textbf{(a)-(b) Asymptotic irrelevance of the initial values.} Results of this type are fairly common in the literature. For a similar argument, see the proof of Theorem~3.1 of \citet{FrancqZakoian2004}. Details can also be found in \citet{Chen2013}.

\textbf{(a) Asymptotic irrelevance of the initial values - I.} In view of the matrix representations of ARMA and GARCH, assumptions \textbf{(A.4)} and \textbf{(B.2)} -- \textbf{(B.4)} imply that almost surely
\begin{align}
\label{Eq:armagarchinitialbound1}
	\sup_{\boldsymbol{\theta} \in \Theta'} |\tilde{\eta}_t(\boldsymbol{\theta}) - \eta_t(\boldsymbol{\theta})| & \le K \rho^t, \ \forall t \in \mathbb{N},\\
\label{Eq:armagarchinitialbound2}
	|\tilde{\sigma}_t^2(\boldsymbol{\theta}) - \sigma_t^2(\boldsymbol{\theta})| & \le K \rho^t \sum_{i=1-r}^{t-1} (|\eta_i(\boldsymbol{\theta})|+1), \ \forall t \in \mathbb{N},
\end{align}
where $K > 0$ and $0 < \rho < 1$ are two generic constants. See also point (a) in the proof of Theorem~\ref{Thm:armaconsistency} for reference.
It then follows that 
\begin{align*}
	\limsup_{n \rightarrow \infty} \sup_{\boldsymbol{\theta} \in \Theta'} D_2^2 (Q_{n,\boldsymbol{\theta}}, \tilde{Q}_{n,\boldsymbol{\theta}}) &\le \limsup_{n \rightarrow \infty} \sup_{\boldsymbol{\theta} \in \Theta'} \frac{1}{n}\sum_{t=1}^n \left| \frac{\eta_t(\boldsymbol{\theta})}{\sigma_t(\boldsymbol{\theta})} -  \frac{\tilde{\eta}_t(\boldsymbol{\theta})}{\tilde{\sigma}_t(\boldsymbol{\theta})}\right|^2 \\
	&=  \limsup_{n \rightarrow \infty} \sup_{\boldsymbol{\theta} \in \Theta'}\frac{1}{n}\sum_{t=1}^n \left| \frac{\eta_t(\boldsymbol{\theta})}{\sigma_t(\boldsymbol{\theta})} -   \frac{\eta_t(\boldsymbol{\theta})}{\tilde{\sigma}_t(\boldsymbol{\theta})} +  \frac{\eta_t(\boldsymbol{\theta})}{\tilde{\sigma}_t(\boldsymbol{\theta})} - \frac{\tilde{\eta}_t(\boldsymbol{\theta})}{\tilde{\sigma}_t(\boldsymbol{\theta})}\right|^2 \\
	&\le \limsup_{n \rightarrow \infty} \sup_{\boldsymbol{\theta} \in \Theta'} \frac{2}{n}\sum_{t=1}^n \left\{\frac{\eta_t^2(\boldsymbol{\theta}) \left|\sigma_t^2(\boldsymbol{\theta}) - \tilde{\sigma}_t^2(\boldsymbol{\theta})\right|}{\sigma_t^2(\boldsymbol{\theta}) \tilde{\sigma}_t^2(\boldsymbol{\theta})} + \frac{(\eta_t(\boldsymbol{\theta}) - \tilde{\eta}_t(\boldsymbol{\theta}))^2}{\tilde{\sigma}_t^2(\boldsymbol{\theta})} \right\}\\
	&\le \limsup_{n \rightarrow \infty} \sup_{\boldsymbol{\theta} \in \Theta'} \frac{2}{n}\sum_{t=1}^n \eta_t^2(\boldsymbol{\theta}) \left|\sigma_t^2(\boldsymbol{\theta}) - \tilde{\sigma}_t^2(\boldsymbol{\theta})\right| \\
	&+ \limsup_{n \rightarrow \infty} \sup_{\boldsymbol{\theta} \in \Theta'} \frac{2}{n} \sum_{t=1}^n (\eta_t(\boldsymbol{\theta}) - \tilde{\eta}_t(\boldsymbol{\theta}))^2 .
\end{align*}
Here we used the fact that $\boldsymbol{\theta} \in \Theta'$, so both $\tilde{\sigma}_t^2(\boldsymbol{\theta})$ and $\sigma_t^2(\boldsymbol{\theta})$ are greater than or equal to one. For the first term, we can apply (\ref{Eq:armagarchinitialbound2}) and a similar argument in the proof of Theorem~3.1 of \citet{FrancqZakoian2004} to prove that it approaches zero almost surely. For the second term, (\ref{Eq:armagarchinitialbound1}) entails its almost sure convergence to zero.

\textbf{(b) Asymptotic irrelevance of the initial values - II.} Utilizing the inequality $|\log x - \log y| \le \frac{|x-y|}{\min(x,y)}$ for $x, y > 0$ and (\ref{Eq:armagarchinitialbound2}), one has that almost surely
\begin{align*}
	\limsup_{n \rightarrow \infty} \sup_{\boldsymbol{\theta} \in \Theta'} \frac{1}{2n} \left|\sum_{t=1}^n \log \tilde{\sigma}_t^2(\boldsymbol{\theta}) - \sum_{t=1}^n \log \sigma_t^2(\boldsymbol{\theta}) \right| &\le \limsup_{n \rightarrow \infty} \sup_{\boldsymbol{\theta} \in \Theta'} \frac{1}{2n} \sum_{t=1}^n \left| \tilde{\sigma}_t^2(\boldsymbol{\theta}) - \sigma_t^2(\boldsymbol{\theta})\right| \\
	&\le \limsup_{n \rightarrow \infty} \sup_{\boldsymbol{\theta} \in \Theta'} \frac{K}{2n} \sum_{t=1}^n \rho^t \sum_{i=1-r}^{t-1} (|\eta_i(\boldsymbol{\theta})|+1).
\end{align*}
The rest of the proof is similar to that of point (a). 

\textbf{(c) Existence of the logarithmic expectation over $\Theta'$.} Here the ARCH($\infty$) representation of GARCH is used. Jensen's inequality and the subadditivity of the function $f(z) = z^u$, $z \in (0,\infty)$ entail that for any $\boldsymbol{\theta} \in \Theta'$,
\begin{align*}
	\mathbb{E} |\log \mathring{\sigma}_1^2(\boldsymbol{\theta})| = \mathbb{E} \log \mathring{\sigma}_1^2(\boldsymbol{\theta}) 
	&\le \frac{1}{u} \log \mathbb{E} \left(\frac{1}{\mathcal{B}_{\boldsymbol{\theta}}(1)} + \sum_{i = 1}^{\infty} \gamma_i(\boldsymbol{\theta}) \mathring{\eta}_{1-i}^2(\boldsymbol{\theta})\right)^u \\
	&\le \frac{1}{u} \log \left( \mathcal{B}^{-u}_{\boldsymbol{\theta}}(1) + \mathbb{E} \mathring{\eta}_{1}^{2u}(\boldsymbol{\theta}) \sum_{i=1}^{\infty} |\gamma_i(\boldsymbol{\theta})|^u  \right),
\end{align*}
where $\{\gamma_i(\boldsymbol{\theta})\}_{i=1}^\infty$ are given as
\[
	\gamma_i(\boldsymbol{\theta}) = \frac{1}{i!}\frac{d^i}{dz^i}\left\{\left.\frac{\mathcal{A}_{\boldsymbol{\theta}}(z)}{\mathcal{B}_{\boldsymbol{\theta}}(z)}\right\}\right|_{z=0}, \mbox{ for } i = 1, 2, \ldots.
\]
Now because all the roots of $\mathcal{B}_{\boldsymbol{\theta}}(z) = 0$ have modulus greater than one and $\Theta'$ is compact, we can find two constants $K > 0$ and $0 < \rho < 1$ such that $\sup_{\boldsymbol{\theta} \in \Theta'}|\gamma_i(\boldsymbol{\theta})| < K \rho^i$ for every $i \in \mathbb{N}$. It therefore follows that $\sup_{\boldsymbol{\theta} \in \Theta'} \sum_{i=1}^{\infty} |\gamma_i(\boldsymbol{\theta})|^u  < \frac{K}{1-\rho^u} < \infty$. 

From Proposition~1 of \citet{FrancqZakoian2004}, there exists an $u \in (0,1/2)$ with $\mathbb{E}\mathring{\eta}_t^{2u}(\boldsymbol{\theta}_0) < \infty$. Using essentially the same argument on the MA$(\infty)$/AR($\infty$) representation of ARMA, we obtain that $\mathbb{E}\mathring{X}_1^{2u} < \infty$ and $\sup_{\boldsymbol{\theta} \in \Theta'}\mathbb{E}(\mathring{\eta}_1^{2u}(\boldsymbol{\theta})) < \infty$. Therefore, $\mathbb{E} |\log \mathring{\sigma}_1^2(\boldsymbol{\theta})|$ is bounded over $\Theta'$.

\textbf{(d) The lower bound.} 
It is easy to check that $Q_{n,\boldsymbol{\theta}_0'} = \frac{1}{n}\sum_{t=1}^n \delta_{\sqrt{c_0} \epsilon_t}$.
Denote by $Q_{0'}$ the distribution corresponding to $\sqrt{c_0}\epsilon_t$. Then $D_1(Q_{n,\boldsymbol{\theta}_0'},Q_{0'}) \stackrel{a.s.}{\rightarrow} 0$. By combining this with point (a), we deduce $D_1(\tilde{Q}_{n,\boldsymbol{\theta}_0'},Q_{0'}) \stackrel{a.s.}{\rightarrow} 0$. Now use point (b), (c) and the pointwise ergodic theorem to see 
\[
	\lim_{n \rightarrow \infty} \frac{1}{2n} \sum_{t=1}^n \log \tilde{\sigma}_t^2(\boldsymbol{\theta}_0') = \frac{1}{2}\,  \mathbb{E} \log \mathring{\sigma}_1^2(\boldsymbol{\theta}_0'), \mbox{ a.s.}
\]
We recall the definition of $\Lambda_n(\cdot,\cdot)$ in (\ref{Eq:armagarchmle1}). 
It then follows from the continuity and the affine equivariance of $L(\cdot)$ (Lemma~\ref{Lem:DSSCont}(ii) and Lemma~\ref{Lem:DSSProp}(ii)) that
\begin{align*}
	\liminf_{n \rightarrow \infty} \sup_{\Phi \times \Theta'} \Lambda_n(\phi, \boldsymbol{\theta}) \ge \liminf_{n \rightarrow \infty} \sup_{\phi \in \Phi} \Lambda_n(\phi, \boldsymbol{\theta}_0') 
	& = \liminf_{n \rightarrow \infty} L(\tilde{Q}_{n,\boldsymbol{\theta}_0'}) - \limsup_{n \rightarrow \infty} \frac{1}{2n} \sum_{t=1}^n \log \tilde{\sigma}_t^2(\boldsymbol{\theta}_0') \\
	& = L(Q_{0'}) - \frac{1}{2} \mathbb{E} \log \mathring{\sigma}_1^2(\boldsymbol{\theta}_0') 
	  = L(Q_0) - \frac{1}{2} \mathbb{E} \log \mathring{\sigma}_1^2(\boldsymbol{\theta}_0). 
\end{align*}

\textbf{(e) Uniform ergodic theorem.} Its proof follows from that of the uniform law of large numbers, where one combines a standard bracketing idea with the compactness argument. We omitted the proof of this part for brevity.

\textbf{(f) Uniform weak convergence.} One may refer to point (c) in the proof of Theorem~\ref{Thm:armaconsistency} for more details, where a similar result has been established.

\textbf{(g) Convergence of $\hat{\boldsymbol{\theta}}_n'$.} To verify the assertion, it suffices to consider a sequence of fixed observations $X_1,X_2, \ldots$ such that (a) -- (f) hold true. Consider any convergent subsequence of $\hat{\boldsymbol{\theta}}_n'$, denoting which by $\hat{\boldsymbol{\theta}}_{n(j)}' \rightarrow \boldsymbol{\theta}^*$. our aim is to show that $\boldsymbol{\theta}^* = \boldsymbol{\theta}_0'$. First, by compactness, $\boldsymbol{\theta}^* \in \Theta'$. Now a slight variant of point (f) together with point (a) entails that
\[
	\lim_{j \rightarrow \infty} D_L(\tilde{Q}_{n(j),\hat{\boldsymbol{\theta}}_{n(j)}'},Q_{\boldsymbol{\theta}^*}) = 0.
\]
For all $\boldsymbol{\theta} \in \Theta'$, 
\[
	\frac{\mathring{\eta}_1(\boldsymbol{\theta})}{\mathring{\sigma}_1(\boldsymbol{\theta})} = \left(\frac{\mathring{\eta}_1(\boldsymbol{\theta}_0')}{\mathring{\sigma}_1(\boldsymbol{\theta}_0')} + \frac{\mathring{\eta}_1(\boldsymbol{\theta}) - \mathring{\eta}_1(\boldsymbol{\theta}_0')}{\mathring{\sigma}_1(\boldsymbol{\theta}_0')}\right) \frac{\mathring{\sigma}_1(\boldsymbol{\theta}_0')}{\mathring{\sigma}_1(\boldsymbol{\theta})} \equiv (R_1 + R_2) R_3,
\]
where $R_1$ is independent of both $R_2$ and $R_3$. So by Lemma~\ref{Lem:DSSProp}(ii), Lemma~\ref{Lem:DSSCont} and Corollary~\ref{Cor:addtimesnoisel}, 
\begin{align}
\label{Eq:armagarchupperbound}
	\limsup_{j \rightarrow \infty} L(\tilde{Q}_{n(j),\hat{\boldsymbol{\theta}}_{n(j)}'}) \le L(Q_{\boldsymbol{\theta}^*}) \le L(Q_0) - \frac{1}{2}\log c_0 - \mathbb{E} \log \mathring{\sigma}_1^2(\boldsymbol{\theta}_0') + \mathbb{E} \log \mathring{\sigma}_1^2(\boldsymbol{\theta}^*).
\end{align}
Furthermore, it is easy to check from points (b) and (e) that
\[
	\lim_{j \rightarrow \infty} \frac{1}{n} \sum_{t=1}^n \log \tilde{\sigma}_t^2(\hat{\boldsymbol{\theta}}_{n(j)}') = \mathbb{E} \log \mathring{\sigma}_1^2(\boldsymbol{\theta}^*).
\]
Combining those two elements together gives that
\begin{align*}
	\nonumber \limsup_{j \rightarrow \infty} \sup_{\Phi \times \Theta'} \Lambda_{n(j)}(\phi, \boldsymbol{\theta}) & \le \limsup_{k \rightarrow \infty} L(\tilde{Q}_{n(j),\hat{\boldsymbol{\theta}}_{n(j)}'}) - \liminf_{j \rightarrow \infty} \frac{1}{2n} \sum_{t=1}^{n} \log \sigma_t^2(\hat{\boldsymbol{\theta}}_{n(j)}') \\
	& \le L(Q_0) - \frac{1}{2}\log c_0 - \frac{1}{2} \mathbb{E} \log \mathring{\sigma}_1^2(\boldsymbol{\theta}_0') = L(Q_0) - \mathbb{E} \log \mathring{\sigma}_1^2(\boldsymbol{\theta}_0).
\end{align*}
In light of point (d), the equality is enforced in (\ref{Eq:armagarchupperbound}). So by Corollary~\ref{Cor:addtimesnoisel} again, there must exist constants $C_1$ and $C_2  \in (0, \infty) $ such that 
\begin{align}
\label{Eq:armagarchidentifiable1}
	 \mathbb{P}\left(\frac{\mathring{\eta}_1(\boldsymbol{\theta}^*) - \mathring{\eta}_1(\boldsymbol{\theta}_0')}{\mathring{\sigma}_1(\boldsymbol{\theta}_0')} = C_1 \right) &= 1, \\
\label{Eq:armagarchidentifiable2}
	\mathbb{P}\left(\frac{\mathring{\sigma}_1^2(\boldsymbol{\theta}_0')}{\mathring{\sigma}_1^2(\boldsymbol{\theta}^*)} = C_2 \right) &= 1 .
\end{align}

Note that for every $\boldsymbol{\theta} \in \Theta'$, one can express $\mathring{\eta}_1(\boldsymbol{\theta})$ as a linear combination of $\mathring{X}_{1-i}, i \ge 0$. Furthermore, one can write $\mathring{\sigma}_1^2(\boldsymbol{\theta}) - 1/\mathcal{B}_{\boldsymbol{\theta}}(1)$ as a linear combination of $\mathring{X}_{1-i}\mathring{X}_{1-j},i,j \ge 1$.  We claim that $C_1 = 0$ and $\mathring{\eta}_1(\boldsymbol{\theta}^*) = \mathring{\eta}_1(\boldsymbol{\theta}_0')$ with probability one, because otherwise (\ref{Eq:armagarchidentifiable1}) would imply the existence of a constant linear combination of $\mathring{X}_{1-i}\mathring{X}_{1-j}$ with $i,j \ge 1$, which would violate assumption \textbf{(B.1)} (or even \textbf{(B.1*)}). By the same argument given in the proof of Theorem~\ref{Thm:armaconsistency}, we get $\mathbi{A}_{\boldsymbol{\theta}^*} = \mathbi{A}_{\boldsymbol{\theta}_0'}$ and $\mathbi{B}_{\boldsymbol{\theta}^*} = \mathbi{B}_{\boldsymbol{\theta}_0'}$.

Moreover, it follows from (\ref{Eq:armagarchidentifiable1}) and (\ref{Eq:armagarchidentifiable2}) that with probability one
\[
	\left\{\frac{C_2\mathcal{A}_{\boldsymbol{\theta}^*}(B)}{\mathcal{B}_{\boldsymbol{\theta}^*}(B)} - \frac{\mathcal{A}_{\boldsymbol{\theta}_0'}(B)}{\mathcal{B}_{\boldsymbol{\theta}_0'}(B)} \right\} \mathring{\eta}^2_1(\boldsymbol{\theta}_0') = \frac{1}{\mathcal{B}_{\boldsymbol{\theta}_0'}(1)} - \frac{C_2}{\mathcal{B}_{\boldsymbol{\theta}^*}(1)}.
\]
It can be seen that this equality holds if and only if 
\[
	\frac{C_2\mathcal{A}_{\boldsymbol{\theta}^*}(z)}{\mathcal{B}_{\boldsymbol{\theta}^*}(z)} = \frac{\mathcal{A}_{\boldsymbol{\theta}_0'}(z)}{\mathcal{B}_{\boldsymbol{\theta}_0'}(z)}, \ \forall |z| \le 1 \quad \mbox{ and } \quad \frac{1}{\mathcal{B}_{\boldsymbol{\theta}_0'}(1)} = \frac{C_2}{\mathcal{B}_{\boldsymbol{\theta}^*}(1)}.
\]
Under assumption \textbf{(B.5)}, it implies $\mathcal{B}_{\boldsymbol{\theta}^*} = \mathcal{B}_{\boldsymbol{\theta}_0'}$, which consequently entails $C_2 = 1$ and $\mathcal{A}_{\boldsymbol{\theta}^*} = \mathcal{A}_{\boldsymbol{\theta}_0'}$. 

Therefore, $\boldsymbol{\theta}^* = \boldsymbol{\theta}_0'$. Finally, since $\Theta'$ is compact and the convergent subsequence is picked arbitrarily, $\hat{\boldsymbol{\theta}}_n' \rightarrow \boldsymbol{\theta}_0'$, as desired.

\textbf{(h) Convergence of $\hat{c}_n$.} In view of point (a), it suffices to show $\mu_2'(Q_{n,\hat{\boldsymbol{\theta}}_n'}) \stackrel{a.s.}{\rightarrow} c_0$. One can follow a similar argument used for point (e) in the proof of Theorem~\ref{Thm:armaconsistency} to establish this point. Moreover, by the continuous mapping theorem, $\hat{\boldsymbol{\theta}}_n \stackrel{a.s.}{\rightarrow}\boldsymbol{\theta}_0$.

\textbf{(i) Convergence of $\hat{f}_n$.} A close scrutiny reveals that we have already established firstly the convergence of $Q_{n,\hat{\boldsymbol{\theta}}_n'}$ to $Q_{0'}$ in law in the proof of point (g), and secondly, $\mu_2'(Q_{n,\hat{\boldsymbol{\theta}}_n'}) \stackrel{a.s.}{\rightarrow} \mu_2'(Q_{0'})$ in point (h). The convergence of $Q_{n,\hat{\boldsymbol{\theta}}_n'}$ to $Q_{0'}$ in the $2^{\mathrm{nd}}$ Mallows distance then follows from Theorem~6.9 of \citet{Villani2009}, which also implies the convergence in the $1^{\mathrm{st}}$ Mallows distance. Again by point~(a), $D_1(\tilde{Q}_{n,\hat{\boldsymbol{\theta}}_n'}, Q_{0'}) \stackrel{a.s.}{\rightarrow} 0$. Now one can use Lemma~\ref{Lem:d1conv}(i) and Lemma~\ref{Lem:DSSCont}(ii) to obtain $\int \bigl| \hat{f}_n(x) - f_0^*(x) \bigr| \ dx \stackrel{a.s.}{\rightarrow} 0$. Finally, one can apply Proposition~2 of \citet{CuleSamworth2010} and the dominated convergence theorem to see (\ref{Eq:armagarchvariance}).
\hfill $\Box$
\end{prooftitle}

\end{document}